\documentclass[%
aip,
jmp,%
amsmath,amssymb, twocolumn,
reprint%
]{revtex4}

\usepackage{graphicx}
\usepackage{dcolumn}
\usepackage{bm}
\usepackage{xcolor}
\usepackage{verbatim}
\usepackage{amsfonts}
\usepackage{amsthm}
\usepackage{bookmark}
\usepackage{braket}
\usepackage{placeins}
\usepackage{subeqnarray}
\graphicspath{{Figures/}}
\usepackage[font=small,labelfont=bf]{caption}
\usepackage{soul}
\usepackage{svg}
\renewcommand\bra[1]{{\langle{#1}|}}
\makeatletter
\renewcommand\ket[1]{%
  \@ifnextchar\bra{\k@t{#1}\!}{\k@t{#1}}%
}
\newcommand\k@t[1]{{|{#1}\rangle}}
\makeatother

\begin{document}
\title[Quantum Darwinism and non-Markovianity in a model of quantum 
harmonic oscillators]{Quantum Darwinism and Non-Markovianity in a Model of Quantum 
Harmonic Oscillators}

\author{S. M. Oliveira}
 \affiliation{Departamento de Fisica, Instituto de Ciências Exatas, Universidade Federal de Minas Gerais, 30123-970 Belo Horizonte, Minas Gerais, 
Brazil.}%Lines break automatically or can be forced with \\
 \email{sheilla@fisica.ufmg.br}
 \author{A. L. de Paula Jr.}%
\affiliation{ 
Departamento de Fisica, Instituto de Ciências Exatas, Universidade Federal de Minas Gerais, 30123-970 Belo Horizonte, Minas Gerais, Brazil%\\This line break forced with \textbackslash\textbackslash
}%
\affiliation{ 
Instituto Federal de Educação, Ciência e Tecnologia do Rio de Janeiro, 20270-021 Rio de Janeiro, Rio de Janeiro, Brazil%\\This line break forced with \textbackslash\textbackslash
}%

\author{R. C. Drumond}
\affiliation{%
Departamento de Matematica, Instituto de Ciências Exatas, Universidade Federal de Minas Gerais, 30123-970 Belo Horizonte, MG, Brazil
}%

\date{\today}% It is always \today, today,

\begin{abstract}
To explain aspects of the quantum-to-classical transition, quantum Darwinism explores
the fact that, due to interactions between a quantum open system and its surrounding environment, information about the system can be spread 
redundantly to the environment. Here we recall that there are in the literature two distinct and non-equivalent ways to make this 
statement precise and quantitative. We first point out the difference with some simple but illustrative examples. We then consider 
a model where Darwinism can be seen from both perspectives. Moreover, the non-Markovianity of our model can be varied with a 
parameter. In a recent work [F. Galve \textit{et al.}, Sci. Reps. 6, 19607 (2016)], the authors concluded that quantum Darwinism can be hindered by 
non-Markovianity. We depart from their analysis and argue that, from both perspectives to quantum Darwinism, there is no 
clear relationship between non-Markovianity and quantum Darwinism in our model.

\end{abstract}

\pacs{Valid PACS appear here}% PACS, the Physics and Astronomy
\keywords{Suggested keywords}%Use showkeys class option if keyword
\maketitle

\section{\label{sec:level1}Introduction}

Since the formulation of quantum theory, problems with the quantum-to-classical transition have been highlighted, for instance, by 
 the Schr\"odinger's cat \emph{gedanken} experiment~\cite{schcat} and the EPR ``paradox''~\cite{epr}. In particular, the 
superposition principle and the quantum state ``fragility'' under measurements are, at a first glance, difficult to conflate with 
basic notions of classical physics. However, considerable progress was made in the past decades by taking into account that effectively 
classical systems, besides being macroscopic, are typically open~\cite{zur1981, zur2003, schl2007}. 

The fact that macroscopic systems are never seen in superposition states with distinct macroscopic properties is essentially explained 
by the effect of decoherence~\cite{zur2003, zur1982, bru1996, dav1996, bre2002, zur2005a, schl2007}. Indeed, every physical system 
constantly interacts with its surrounding environment. For macroscopic systems this interaction can 
never be completely neglected (except, perhaps, under extremely artificial laboratory conditions). In fact, it implies that any coherence between macroscopically 
distinct states will quickly vanish, transforming superpositions of such states into statistical mixtures. 

Openness is also an important ingredient to explain why macroscopic systems, when in their naturally realized states, are essentially 
insensitive to measurements. In Refs.~\cite{zur2004, zur2003a, zur2005, zur2006, zur2005b, zunatk2009} the authors observed that the 
environment can monitor a so-called preferred observable of the system and store information about it. This is done in such a way that it is possible to 
extract information about it indirectly and disturbing the system minimally (beyond what it was already disturbed by the environment). More 
than that, different regions of the environment record that information redundantly, so it is sufficient to access just a small fragment to 
obtain a significant amount of information about the preferred observable of the macroscopic system. This is roughly the idea behind quantum 
Darwinism.

The concept of quantum Darwinism has been explored in several 
models, like in spin 
systems \cite{zur2009}, where the environment monitors the spin of a particle; in a quantum Brownian particle 
\cite{zur2008}, where the environment monitors its position; and recently in an experimental work, within a nitrogen vacancy center 
\cite{zur2018}. Moreover, the role of Markovianity in the emergence of quantum Darwinism was studied recently in Refs.~\cite{mas2019} 
and \cite{sab2016}. In Ref.~\cite{sab2016} the authors discuss a relationship between quantum Darwinism and non-Markovianity concluding 
that non-Markovianity can hinder quantum Darwinism.

As originally proposed (see, for instance, Ref.~\cite{zur2003}), from the global 
system-environment state at some instant of time, 
one can compute the quantum mutual information between the system and fractions of the 
environment. That can be used to measure both the 
amount and redundancy of information about the system state that is available in the environment \emph{at 
that instant of time}, regardless of the initial state of both system and environment. Alternatively, 
one can look at how initial states of the system are mapped to states in 
fractions of the environment after some interaction time~\cite{bra2015, hor2015}. From that, one can check 
if some information about a system observable 
\emph{before} the interaction can be recovered from measurements in these environment fractions 
\emph{after} the interaction.

In this paper we address the differences in these two approaches, first, through simple examples. We then investigate them in a 
model of a quantum harmonic oscillator coupled to an environment also composed of quantum harmonic oscillators. Furthermore, we can range 
the dynamics of the main oscillator from Markovian to non-Markovian by tuning one of the parameters of the model. This allows us to explore 
carefully the relationship between non-Markovianity and quantum Darwinism. Following the definition given 
in~\cite{sab2016} to estimate the ``degree of quantum Darwinism'', we get results qualitatively similar to theirs, in the sense that it 
suggests that ``non-Markovianity hinders quantum Darwinism''. However, we argue that we should estimate quantum Darwinism from other 
points of views. From these, we actually do not see any clear relationship between the two concepts.

In Sec.~\ref{QDT} we recall the ideas behind two approaches to quantum Darwinism and, in Sec.~\ref{NonM}, 
some basic notions of non-Markovianity. We define the model in Sec.~\ref{model} and discuss its relevant 
properties. In Sec.~\ref{resultsPIP} and \ref{resultsBPH} we discuss quantum Darwinism in our model from these two aforementioned 
perspectives. In Sec.~\ref{darwinismxmarkovianity} we discuss the relationship (or absence of it) between quantum Darwinism and 
non-Markovianity. Finally, we close our paper with some concluding remarks in Sec.~\ref{conclusion}.

\section{Quantum Darwinism(s)}\label{QDT}

Whenever a system $S$ and its environment $E$ interact with each other, information about the system can be transferred to the environment 
and, in some cases, the environment behaves, effectively, as a measurement apparatus \cite{zur1982, 
zur2003}. Typically, and especially for macroscopic systems, superposition states among elements in some special basis are rapidly 
``destroyed'' and they lose coherence. This process is called decoherence~\cite{bre2002} and that basis is referred to as the preferred 
basis (or observable) of the system.

The main idea of quantum Darwinism is that one additionally should take into account that the environment is composed of several 
subsystems. One can consider then distinct fragments composed by a subset of these subsystems. These fragments can record information 
redundantly about the macroscopic system in such a way that just a small piece of the environment is enough to 
obtain nearly all information available in the whole environment about the 
preferred observable of the system.

For a given model in which Darwinism is expected to take place, it is important to identify the system observable that the 
environment monitors. For instance, in the model studied in Ref. \cite{zur2008}, the preferred 
observable is the position of a quantum harmonic oscillator, while in the model studied in 
Ref. \cite{zur2009} the observable monitored is the $z$ component of a spin-$1/2$ system. 

The monitored observable is easily recognized when, for some orthonormal basis $\{\ket{\phi^i}\}$ in the system 
Hilbert space and an environment initial state $\ket{\Psi}$, the global dynamics is
\begin{equation}
(\sum_{i}\alpha_{i}\ket{\phi^i})\otimes \ket{\Psi}\mapsto \sum_{i}\alpha_{i}\ket{\phi^i}\otimes \ket{\Psi^{i}}, 
\label{exemplodarwinismo}
\end{equation}
where $\alpha_{i}$ are the complex coefficients of the system initial state and $\ket{\Psi^{i}}$ are 
environment states after the interaction with the system. If the states $\ket{\Psi^{i}}$ are mutually orthogonal, a global measurement 
on the environment, in a basis including the vectors $\ket{\Psi^{i}}$, is equivalent to a measurement on the system in the basis 
$\{\ket{\phi^i}\}$, making the former the preferred basis. Note that, in this case, evolution Eq.~\eqref{exemplodarwinismo} is exactly the 
premeasurement of the von Neumann measurement scheme~\cite{von2018mathematical}. In the Darwinist regime, however, already local 
measurements on small portions of the environment should be enough. An extreme case would be the one where 
$\ket{\Psi^{i}}=\otimes_{k=1}^{N}\ket{\psi_{k}^{i}}$, if the environment has $N$ individual subsystems, with $\{\ket{\psi_{k}^{i}}\}$ being 
mutually orthogonal states of an individual 
environment subsystem $k$. Indeed, a measurement on environment subsystem $k$, in a basis which includes the vectors  
$\ket{\psi_{k}^{i}}$, is already equivalent to a measurement on the system in the basis $\{\ket{\phi^i}\}$. 
 
For a general dynamics, or for general states, one needs a proper framework to address the problem. In the next section, we discuss the two 
main approaches available in the literature.
 
\subsection{The \emph{Partial Information Plot} approach} 

\begin{figure}
\includegraphics[width=0.7\linewidth]{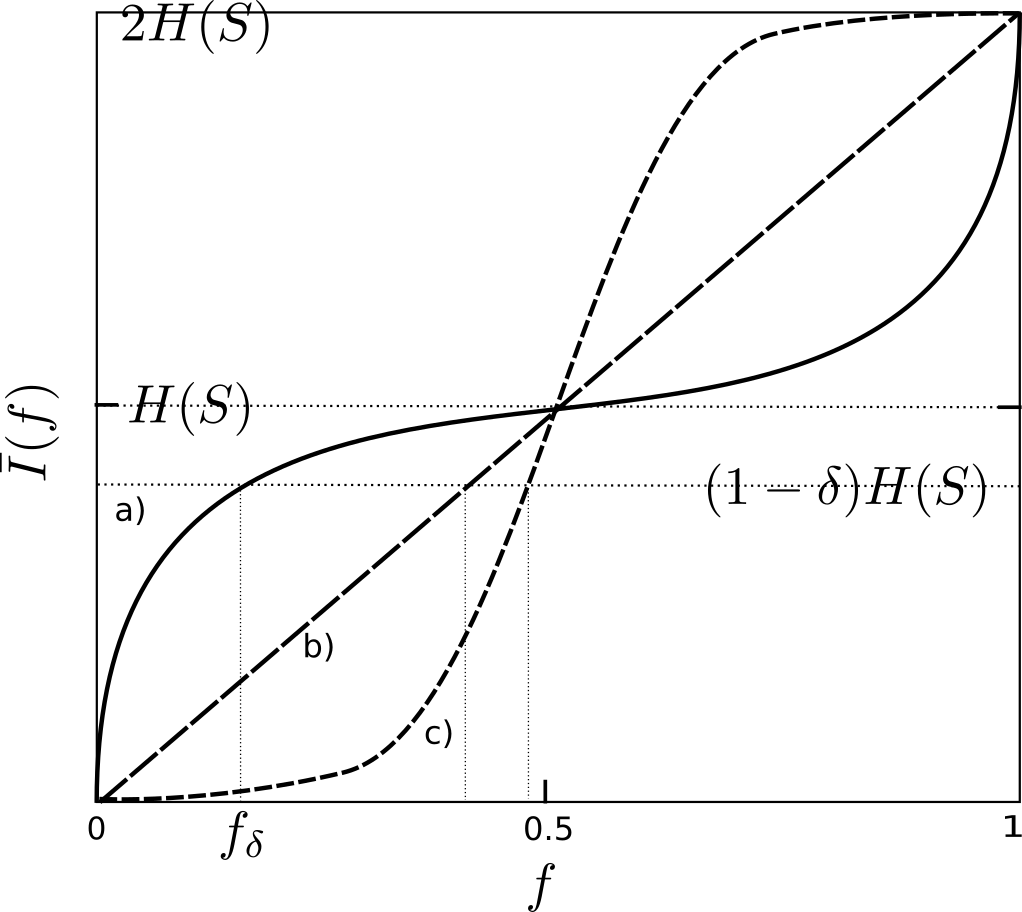}
\captionof{figure}{Three qualitatively distinct possibilities for a partial information plot (PIP). The solid line $(a)$ corresponds 
to a case where a small fraction of the environment already has average mutual information close to $H(S)$, being the signature of  
quantum Darwinism. The linear profile $(b)$ can be seen (approximately) in our model for small interaction times (see Sec.~\ref{resultsPIP}). 
In profile $(c)$, $f_{\delta}$ is close to $0.5$, so the redundancy is very small. This kind of profile can be obtained from random pure 
states drawn according to the Haar measure~\cite{zur2005}. }\label{pipfdelta}  
\end{figure}

As initially proposed by Blume-Kohout and Zurek, the quantum mutual information can be used to quantify the redundancy of information about 
the system that is available in the environment, through the so-called \emph{partial information 
plot} (\emph{PIP})~\cite{zur2005}. Assume a quantum system $S$ and an environment $E$ composed of $N$ individual quantum systems. If the global 
state is $\rho$, for a certain environment fragment $F\subset E$ (that is, any subset of quantum systems from the environment), the mutual 
information between $S$ and $F$ is
\begin{equation}
I(S:F)=H(S)+H(F)-H(SF),\noindent
\label{mi}
\end{equation}
where $H(S)$, $H(F)$ and $H(SF)$ are the von-Neumann entropies of $S$, $F$ 
and $S+F$, respectively, with $H(X)=-\text{Tr}[\rho_{X}\ln{\rho_{X}}]$ for the reduced state $\rho_{X}$ of 
subsystem $X$. The idea is to compute $\bar{I}(f)$, defined by the average 
of $I(S:F)$ over all possible environment fragments $F$ composed by $fN$ individual subsystems, where $0\leq f\leq 1$ is the \emph{fraction} 
of the environment that is taken. 
For pure global states, this curve is always antisymmetric with respect to the coordinate system defined by $\bar{I}=H(S)$ and $f=0.5$~\cite{zur2005} (see Fig.~\ref{pipfdelta}). Moreover, $H(S)$ should be the 
maximum amount of information that can be retrieved about the preferred observable (see Sec.~\ref{examples} for an example). Therefore, in 
the Darwinistic regime, $\bar{I}(f)$ should be close to $H(S)$ already for small $f$, and the PIP should look 
like a plateau (see the solid line in Fig.~\ref{pipfdelta}), since that implies that almost all information about the preferred observable 
is already available in small fractions of the environment. 

Considering the above, it is useful to define, for arbitrary $0<\delta<1$, the smallest fragment size $f_{\delta}$ such that 
$\bar{I}(f_{\delta})>(1-\delta)H(S)$. The quantity \begin{equation}
R_\delta=\frac{1}{f_\delta}\noindent\label{rdelta}
\end{equation}
measures then the \emph{redundancy} in the information about the preferred observable of the system. Indeed, the closer the PIP is to a plateau, the smaller is $f_{\delta}$ for a fixed value of $\delta$ and, hence, the larger $R_{\delta}$ is 
(see Fig.~\ref{pipfdelta}).

\subsection{The \emph{BPH} approach}

An alternative idea is to look at how initial states of the system are mapped into states of the environment, as proposed 
in Ref.~\cite{bra2015} by F. Brand\~ao \textit{et. al.}, which we call the BPH approach for short (see also a related approach in~\cite{hor2015}). There, they take as the 
fundamental object a completely positive map $\Lambda_{E}:D_{S}\rightarrow D_{E}$, where 
$D_{S}$ and $D_{E}$ are the state spaces of the system and environment, respectively. Physically, this map is the result of partial 
tracing the \emph{system} after system and environment evolved for some fixed amount of time and a fixed initial environment state (see 
Sec.~\ref{examples} for an example). Since the 
environment is composed of many subsystems, one can define, via composition with partial tracing, corresponding maps 
$\Lambda_{F}:D_{S}\rightarrow D_{F}$ to each fragment $F$ of the environment, 
namely, $$\Lambda_{F}=\text{Tr}_{E-F}\circ \Lambda_{E}.$$  

A preferred observable would be in general described by a positive operator valued measure (POVM) $\{ 
M_{i}\}$ and distinguished by the condition:
$$\Lambda_{F}(\rho)\approx \sum_{i}\text{Tr}(M_{i}\rho)\sigma_{i,F}$$
for every system operator $\rho$ and some states $\sigma_{i,F}$ for environment fraction $F$. The approximation must hold for \emph{most} 
environment fractions $F$ and the POVM $\{M_{i}\}$ must be independent of $F$. The expression on the right-hand side is a so-called measure 
and prepare map. If states $\sigma_{i,F}$ are sufficiently distinguishable (for distinct values of $i$), some measurement can be done on 
$F$ to infer the statistics of the preferred observable on the state $\rho$, that is, the set of values $\{\text{Tr}(M_{i}\rho)\}$.  

They further formulate the concepts of emergent objectivity of the observables and outcomes and show their validity 
under some circumstances \cite{bra2015, kno2018}. The objectivity of observables guarantees that all observers have access to the same 
observable (namely, $\{M_{i}\}$ is independent of $F$), and it is a generic feature of quantum mechanics~\cite{bra2015}. The objectivity of 
outcomes occurs when different 
observers, who access different fragments of the environment, agree with the results of their measurements. It is guaranteed to take place 
when the outcomes of measurement on these fragments contain almost all information about the preferred observable (roughly speaking, for almost all $F$, the states $\sigma_{F,i}$ 
must be 
highly distinguishable for distinct values of $i$). 

\subsection{Examples}\label{examples}

We discuss here two simple examples with a two-fold objective: to highlight the ideas behind both perspectives of quantum Darwinism, 
making explicit their non-equivalence, and to serve as toy models to two distinct time regimes of the model we shall discuss in 
Sec.~\ref{model}.

\textbf{Example 1.} Assume the main system $S$ is a qubit and the environment $E$ is composed of $N$ qubits, for which we use the set 
of labels $E=\{1,...,N\}$. Assume further that the environment initial state is $\ket{0}_{E}:=\otimes_{k\in E}\ket{0}_{k}$ and their joint 
dynamics, after some fixed time interval $T$, is given by
\begin{align}
\ket{0}\ket{0_E}\mapsto \ket{0}\ket{0_{E}},\\
\ket{1}\ket{0_E}\mapsto \ket{1}\ket{1_{E}},
\end{align}
where $\ket{1_{E}}:=\otimes_{k\in E}\ket{1}_{k}$. Therefore, if the initial state of the system is $\frac{1}{\sqrt{2}}(\ket{0}+\ket{1})$, 
the evolved state is the GHZ state
\begin{equation}
\frac{1}{\sqrt{2}}(\ket{0}\ket{0_{E}}+\ket{1}\ket{1_{E}}).\label{GHZ}
\end{equation}
For any proper fragment $F\subset{E}$ of the environment, it holds:
$$\rho_{SF}=\frac{1}{2}\ket{0}\bra{0}\ket{0_F}\bra{0_{F}}+\frac{1}{2}\ket{1}\bra{1}\ket{1_F}\bra{1_{F}},$$
where $\ket{0_{F}}=\otimes_{k\in F}\ket{0}_{k}$ and analogously for $\ket{1_{F}}$. Therefore, by measuring the environment fraction $F$ in 
the basis $\{\ket{0_F},\ket{1_F}\}$, one can predict with a certainty the outcome of a $\{\ket{0},\ket{1}\}$ measurement in $S$, at 
time $T$. On the other hand, no measurement whatsoever in $F$ will reveal any information about, say, 
a $\{\ket{+},\ket{-}\}$ measurement in $S$, where $\ket{\pm}=\frac{1}{\sqrt{2}}(\ket{0}\pm\ket{1})$. Indeed, it is easy to check that for both outcomes of such a measurement in the system, the corresponding states, for a proper environment fragment, are identical. Therefore, no measurement in a proper fragment would be able to distinguish these two states and there would be no way to predict (better than a random guess) the measurement outcome of the system. This reinforces the fact that $\{\ket{0},\ket{1}\}$ is the preferred basis (or 
observable). 

However, the global state can also be written as
$$\frac{1}{\sqrt{2}}(\ket{+}\ket{\text{GHZ}_{+}}+\ket{-}\ket{\text{GHZ}_{-}}),$$
where $\ket{\text{GHZ}_{\pm}}=\frac{1}{\sqrt{2}}(\ket{0_{E}}\pm\ket{1_{E}}).$ Therefore, if one has access to the whole environment, a measurement in the basis 
$\{\ket{\text{GHZ}_{+}},\ket{\text{GHZ}_{-}}\}$ actually allows one to infer the outcome of a $\{\ket{+},\ket{-}\}$ measurement in $S$. Note that, moreover, the same is true for any system observable, defined by an arbitrary orthonormal basis.

It is easy to check that the PIP for state Eq.~\eqref{GHZ} is an 
exact plateau: $\bar{\mathcal{I}}(0)=0$ by definition, $\bar{\mathcal{I}}(f)=H(S)=\ln{2}$ for every $0<f<1$ and 
$\bar{\mathcal{I}}(f)=2H(S)=2\ln{2}$ for $f=1$. This reflects the fact that any proper subset of the 
environment only has information about the preferred observable. The environment as a whole, on the other hand, has potential information about any observable of the system. But, of course, to predict with certainty the outcomes of other observables, a global measurement in the whole environment must be done (such as a measurement in the basis $\{\ket{\text{GHZ}_{+}},\ket{\text{GHZ}_{-}}\}$), so there is no redundancy whatsoever.

This example can be also understood from the BPH perspective. Here we have, for any environment fragment $F\subset E$,
\begin{equation}
\Lambda_{F}(\rho)=\text{Tr}(\ket{0}\bra{0}\rho)\ket{0_{F}}\bra{0_{F}}+\text{Tr}(\ket{1}\bra{1}\rho)\ket{1_{F}}\bra{1_{F}}.\label{LambdaF}
\end{equation}
That is, the maps are exactly measure and prepare maps, with POVM $\{\ket{0}\bra{0},\ket{1}\bra{1}\}$, and the corresponding 
environment states are perfectly distinguishable. Therefore, we can think that the environment acquires redundant information about the 
preferred observable on the \emph{initial} state of the system. More specifically, a $\{\ket{0_{F}},\ket{1_{F}}\}$ measurement on the environment has exactly the same statistics of a $\{\ket{0},\ket{1}\}$ measurement on $S$ for every system state $\rho$. Moreover, one has complete objectivity of outcomes. If observers measure environment qubits in the basis $\{\ket{0},\ket{1}\}$, they 
will always agree on the outcome. For instance, if some of them observes outcome $0$, all of them will also observe that same outcome. 

Note that in the PIP approach, since one considers only the specific global state Eq.~\eqref{GHZ}, a strong claim can be made: If someone observes a certain outcome upon a $\{\ket{0},\ket{1}\}$ measurement in an  environment fragment $F$, we can be certain that a $\{\ket{0},\ket{1}\}$ measurement in the system itself, if performed, will show the same outcome. 
In the BPH there is no analog claim. It is not correct to say in general that, upon seeing the outcome $0$ in an environment $F$, the system \emph{was} in state $\ket{0}$ at $T=0$, since the system could have been, say, in the initial state $\frac{1}{\sqrt{2}}(\ket{0}+\ket{1})$ and there would be still a positive probability for observing the outcome $0$ in the environment fractions. 

It is interesting to note that, from this perspective, in contrast to the PIP approach, having access to the whole environment does not have any additional consequence. One still is restricted to obtain information about the preferred observable .

\textbf{Example 2.} Now, consider a slight modification in the dynamics:
\begin{align}
\ket{0}\ket{0_E}\mapsto \ket{0}\ket{0_{E}},\\
\ket{1}\ket{0_E}\mapsto \ket{0}\ket{1_{E}}.
\end{align}
No matter what initial state we choose for system $S$, it will never correlate with the environment, so the PIP 
is always trivial. From this perspective then, it appears that this is a bad instance of quantum Darwinism. 

From the BPH perspective, however, they are essentially the same. Indeed, the maps $\Lambda_{F}$ defined by this dynamics are 
exactly the same as in Eq.~\eqref{LambdaF} for every \emph{proper} subset $F\subset E$. Nevertheless, as a side note, it is interesting to see that they do differ for $F=E$. In this case, having access to the whole environment does have a consequence. Indeed, the dynamics essentially transfer any initial state of the system to the environment:
\begin{align}
(\alpha\ket{0}+\beta\ket{1})\ket{0_E}\mapsto \ket{0}(\alpha\ket{0_{E}}+\beta\ket{1_{E}}).\label{exemplo2}
\end{align}
Therefore, one can choose to obtain information about an arbitrary system observable, in the sense that there exists some (in general, global) measurement that can be done in the whole environment that will have the exact same statistics of the corresponding system observable in the initial system state. Of course, these measurements being global makes meaningless the notion of objectivity, since it would not be possible to compare the outcomes of different observers.

\section{Non-Markovianity}\label{NonM}
A quantum dynamical system whose quantum maps $\Lambda_{t,t'}$ are divisible in other completely positive maps, i.e., 
$\Lambda_{t_2,t_0}=\Lambda_{t_2,t_1}\Lambda_{t_1,t_0}$ for all $t_0\le t_1\le t_2$, is said to be \textit{Markovian}. Otherwise, the 
evolution is classified as \textit{non-Markovian} \cite{wol2008,riv2014}. There are several different ways to quantify 
non-Markovianity \cite{riv2014,nad2015, nad2017}, and in this paper we estimate the non-Markovianity 
degree through the time variation of the distance between two random initial states~\cite{lac2015}.

It is known that in a Markovian system, the distinguishability between any two states never increases and the 
flow of information has just one direction: from the system to the environment. Yet, in a non-Markovian dynamics this is not guaranteed. As 
they evolve, they can become more distinguishable and the information received by the environment can go back to the 
system~\cite{sab2011,sab2017}. 

One possible way to measure the distance between two states $\rho_1$ and $\rho_2$ is by the fidelity,
\begin{equation}
\mathcal{F}(\rho_1,\rho_2)= \text{Tr}\sqrt{\sqrt{\rho_1}\rho_2\sqrt{\rho_1}},\label{fid}\noindent
\end{equation}
which behaves monotonically under the action of any quantum map. Then, we can define the non-Markovianity degree as 
\begin{equation}
\mathcal{N}=\underset{\rho_1,\rho_2}{\text{max}}\left[-\int_{\dot{\mathcal{F}}<0}\dot{\mathcal{F}}(\rho_1,
\rho_2)dt\right],\label{deg}\noindent
\end{equation}
where the maximization is over all possible pairs of initial states and $\dot{\mathcal{F}}$ is the time derivative \cite{bre2009}. 

\section{The Model}\label{model}

We now consider a model of a main quantum harmonic oscillator coupled to an environment of several quantum 
harmonic oscillators. The 
Hamiltonian, for $N$ environment oscillators, is defined by
\begin{equation}\label{hamiltonian}
H=\hbar\omega_0 a^{\dagger}a + \hbar\sum_{k=1}^N\omega_k b_k^{\dagger}b_k + \hbar\sum_{k=1}^N\gamma_k(a^\dagger b_k + ab_k^\dagger)\noindent,
\end{equation}
where $\omega_0$ is the main oscillator frequency, $\omega_k$ is the $k$-th environment oscillator frequency, $a$, $a^\dagger$, 
$b_k$, and $b_k^\dagger$ are the creation and annihilation operators of the main system and the $k$-th environment oscillator, and 
$\gamma_k$ are coupling constants for the interaction between the $k$-th environment oscillator and the main oscillator. We shall set from 
now on $\hbar=1$.  

Such model can be used to describe dissipation in optical cavities~\cite{scu1997}, the dynamical 
properties of multipartite entanglement~\cite{bet2014} and principles of quantum thermodynamics~\cite{rap2018}. Here we show that 
it also offers a good benchmark to study quantum Darwinism and 
its connection with Markovianity.

Assume that initially the environment is in the vacuum state and the main oscillator is in a coherent state with parameter 
$\alpha_{0}\in\mathbb{C}$, that is,
\begin{equation}
\ket{\alpha_0}\otimes\prod_{k=1}^N\ket{0_k}.
\end{equation}
The global state evolution can be obtained exactly by the ansatz~\cite{scu1997}
\begin{equation}\label{estado}
\ket{\alpha(t)}\otimes\prod_{k=1}^N\ket{\lambda_k(t)},\noindent
\end{equation}
where $\ket{\alpha(t)}$ and $\ket{\lambda_k(t)}$ denote coherent states of the main oscillator and of the $k$-th environment 
oscillator, respectively. Schr\"odinger's equation with the Hamiltonian Eq.~\eqref{hamiltonian} is satisfied as long as the coherent states parameters $\alpha(t)$ and $\lambda_k(t)$ satisfy 
\begin{eqnarray}
i\dot\alpha(t) &=&\omega_0\alpha(t)+\sum_k\gamma_k\lambda_k(t),\label{alphadiff}\\
i\dot\lambda_k &=&\omega_k\lambda_k+\gamma_k\alpha(t)\label{lamdiff}.\noindent
\end{eqnarray}

We will consider the initial state of the main oscillator in superpositions of coherent states and the environment in the 
vacuum state,
\begin{equation}
\ket{\Psi(0)}=G\left[ \left( a\ket{\alpha_0} + b\ket{-\alpha_{0}} \right) \otimes\prod_{k=1}^N\ket{0_k}\right]\noindent,
\label{cat}
\end{equation}
where $\ket{\pm \alpha_0}$ are coherent states with parameters $\pm\alpha_0$, $a$ and $b$ are complex coefficients, and 
the normalization factor is given by $G=(|a|^2 + |b|^2 + ab^*\braket{-\alpha_0|\alpha_0} +\text{c.c.})^{-1/2}$. 

From the linearity of Schrodinger's equation, the state of the composed system 
for any time $t$ is
\begin{equation}
\ket{\Psi(t)}=G\left( a\ket{\alpha(t)}\otimes\ket{\Lambda(t)}+b\ket{-\alpha(t)}\otimes\ket{-\Lambda(t)} \right)\noindent, 
\label{evolvedstate}
\end{equation}
where 
\begin{equation}
\ket{\pm\Lambda(t)}=\prod_{k=1}^{N}\ket{\pm\lambda_k(t)}\noindent\label{lamchi1},
\end{equation}
while $\alpha(t)$ and $\lambda_k(t)$ are the solutions of Eqs.~\eqref{alphadiff} and \eqref{lamdiff} subjected to the initial conditions 
$\alpha(0)=\alpha_{0}$ and $\lambda_{k}(0)=0$. Note, moreover, that the Hamiltonian conserves the total amount of 
excitations, which implies that $|\alpha(t)|^{2}+\sum_{k\in E}|\lambda_{k}(t)|^{2}=|\alpha_{0}|^{2}$ for all $t$. 

We shall consider two kinds of coupling distributions $\gamma_k$ as a function of $k$. In the first case, all environment oscillators are 
coupled to the main oscillator with the same magnitude. In the second case, all environment oscillators, but one, are coupled with the main 
oscillator with the same magnitude. That is, in general, we take
\begin{equation}\label{gammac}
\gamma_k=\left\{ \begin{matrix} \gamma,& \text{if}\ \omega_k\ne\omega_0\\ \bar{\gamma},& \text{if}\ \omega_k=\omega_0 \end{matrix}\right. \ 
\text{for some constants}\ \gamma,\ \bar\gamma.
\end{equation}

It is possible to obtain analytical results to Eqs.~\eqref{alphadiff} and \eqref{lamdiff} in the limit of a continuum of oscillators in the 
environment. Namely, by  making the substitutions \begin{eqnarray}\label{continuum}
\sum_k\ &\rightarrow&\ \int\rho(\omega)d\omega, \nonumber\\ 
\gamma_k\ &\rightarrow&\ \gamma(\omega),\noindent \label{contlimit}\\ \nonumber
\lambda_k(t)\ &\rightarrow&\ \lambda(\omega,t),\noindent
\end{eqnarray}
where $\rho(\omega)$ is the density of oscillators with frequency $\omega$, that is, $\rho(\omega)d\omega$ is the number of oscillators in 
the environment with 
frequencies in the interval $(\omega,\omega + d\omega)$. Then, $|\lambda(\omega,t)|^2\rho(\omega)$ becomes a density of excitations in 
oscillators with 
frequency $\omega$, for each instant of time $t$. 

The relevant parameter for the dynamic is $\gamma^2(\omega)\rho(\omega)$. If we assume it to be constant,
\begin{equation}
\gamma^2(\omega)\rho(\omega)=\gamma^2\rho=\text{const}=C, \label{defdec}\noindent
\end{equation}
one can show that $|\alpha(t)|^{2}$ decays exponentially at a rate $\Gamma=4\pi C$.  The asymptotic 
density of excitations in the environment, that is, $\lim_{t\rightarrow 
\infty}|\lambda(\omega,t)|^2\rho$, is a Lorentzian of width 
$\Gamma$, centered at the main oscillator frequency $\omega_{0}$. For the non-constant coupling case, and in the regime $\bar{\gamma}\gg 
\Gamma$, $|\alpha(t)|^{2}$ still has an exponential envelope with decaying rate $\Gamma/2$, but it also oscillates at a frequency 
$\Omega\approx \bar{\gamma}$. The asymptotic density of excitations is approximately a superposition of two Lorentzians of width $\Gamma/2$, centered 
at $\omega_{0}\pm \Omega$ (see Appendix~\ref{exato} for details).  

Since we have to sort environment fragments $F$ to approach quantum Darwinism in the model, we actually consider exact 
numerical solutions of Eqs.~\eqref{alphadiff} and \eqref{lamdiff} for a large but finite number of environment oscillators. This was 
done, however, in such a way that these solutions are well approximated by those of the continuum limit. For both cases of coupling 
distributions, we have used $N=900$, $\omega_0=1$, while the environment frequencies were linearly distributed from 
$0.1$ to $1.9$.

\begin{figure}[h!]
 \includegraphics[width=.7\linewidth]{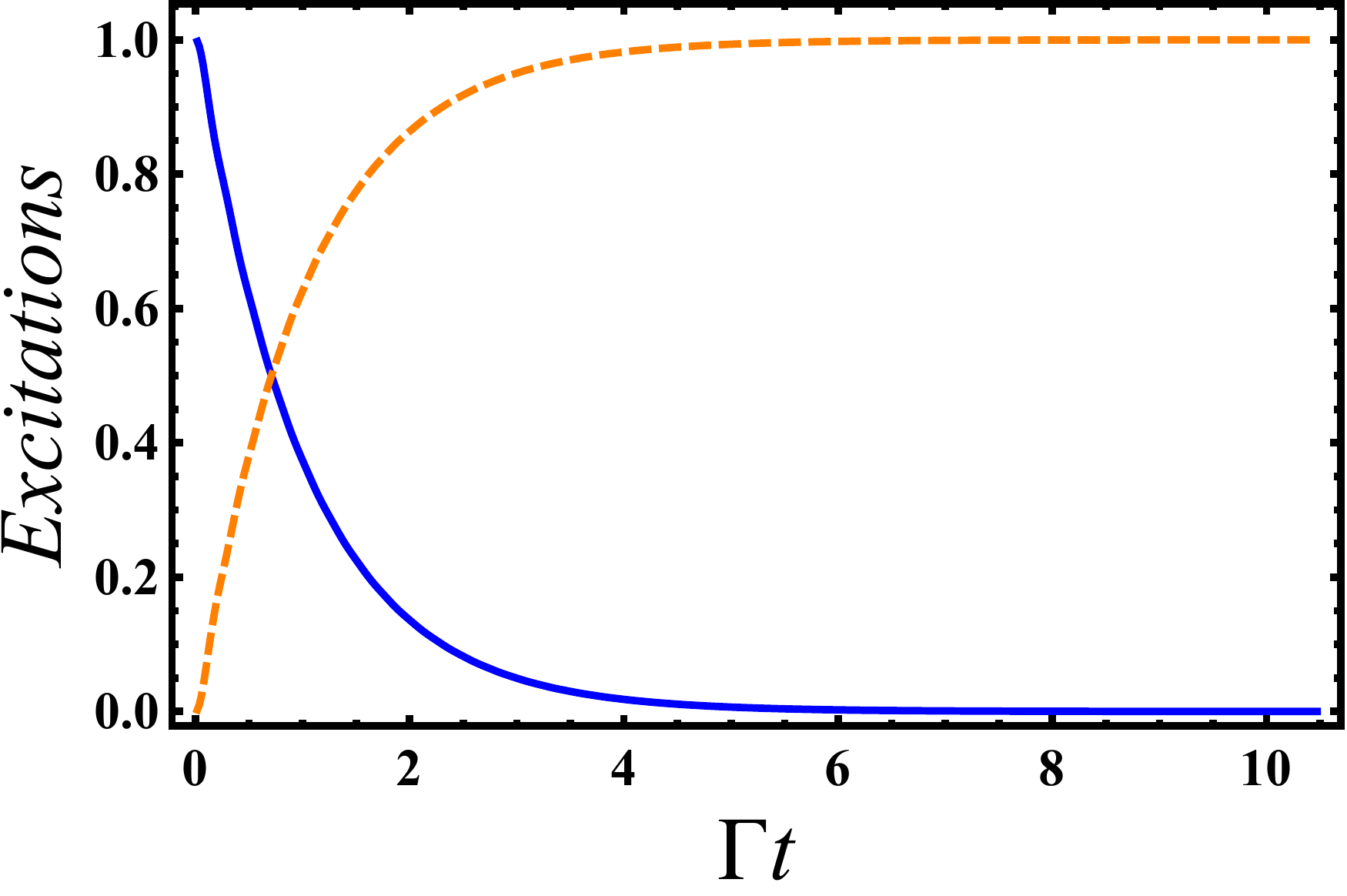}
\captionof{figure}{Dynamics of excitations in the main oscillator $(|\alpha(t)/\alpha_0|)^2$ (blue, solid line) and in the environment $\sum_{k\in E}(|\lambda_{k}(t)/\alpha_0|)^2$ (orange, dashed line) for the constant coupling case $\gamma=\bar{\gamma}=0.1/30$.} \label{conscoupl}  
 \end{figure}

The exact numerical results for the dynamics of excitations in the constant coupling case $\gamma_k=0.1/30$, for all $k$, are shown in 
Fig.~\ref{conscoupl}. While the main system excitations decreases exponentially, as expected, the environment excitations increases to conserve the total amount of excitations. In the long run, almost all 
main system excitations will be transferred to the environment. From the analytical solution we can estimate the decay rate as 
$\Gamma=4\pi\gamma^2\frac{N}{\Delta \omega}$, where $\Delta \omega=1.9-0.1$ is the window of frequencies of environment oscillators and, therefore, $N/\Delta \omega$ is the (uniform) density of oscillators. 
 
\begin{figure}[h!]
 \includegraphics[width=.7\linewidth]{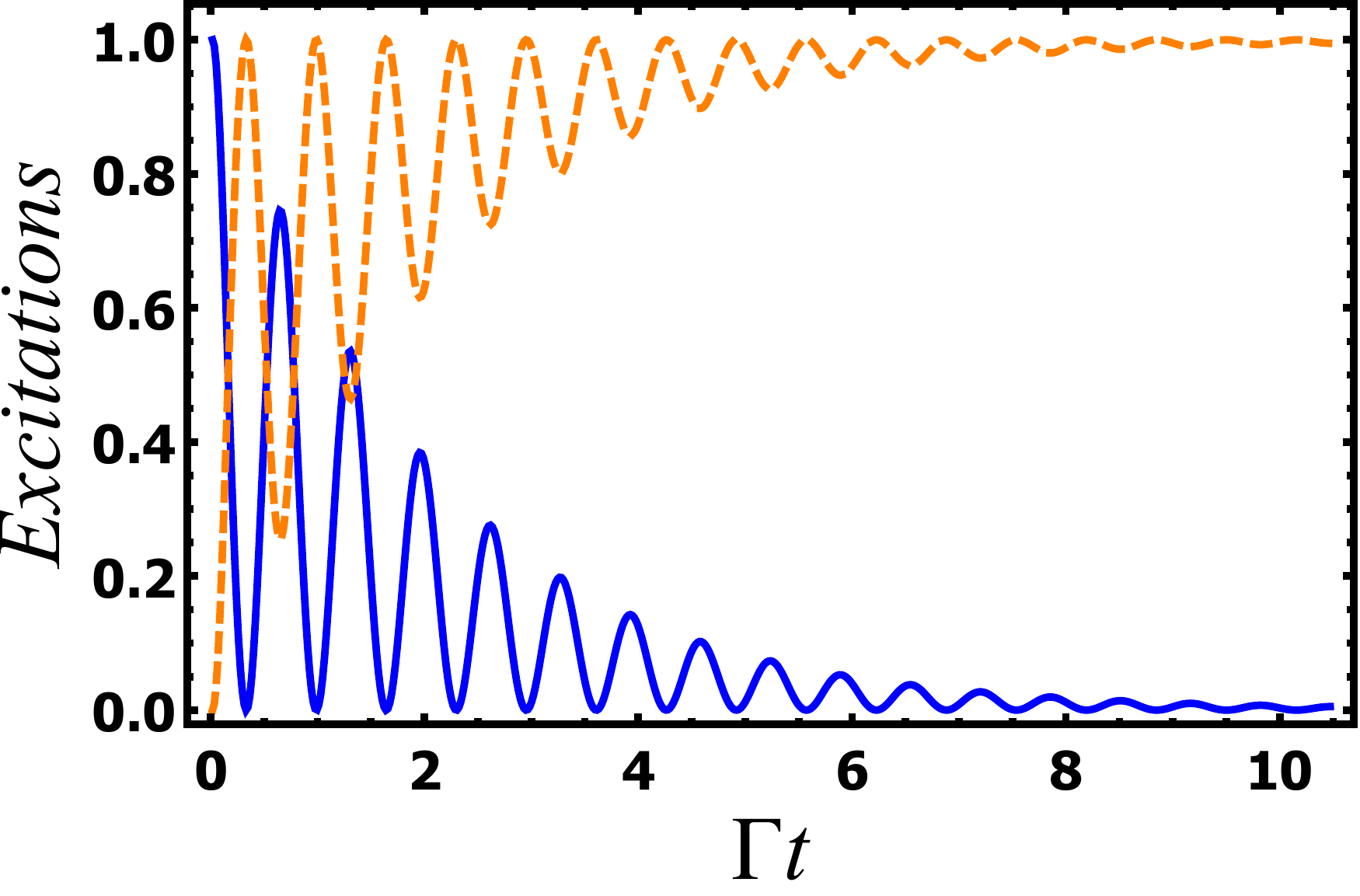}
\captionof{figure}{Dynamics of excitations in the main oscillator $(|\alpha(t)/\alpha_0|)^2$ (blue, solid line) and in the environment $\sum_{k\in E}(|\lambda_{k}(t)/\alpha_0|)^2$ (orange, dashed line) for non-constant couplings, $\gamma=0.1/30$ and $\bar{\gamma}=50\gamma$. 
}\label{lorcoupl50}
 \end{figure}

With a non-constant distribution, namely, $\gamma=0.1/30$ and $\bar{\gamma}=50\gamma$, the transfer of excitations oscillates, where 
the main oscillator is able to take back part of the excitations from the environment (see Fig.~\ref{lorcoupl50}). After a long enough 
time, the amplitude of these oscillations decreases until becoming almost null.  

Observing the excitations behavior, two regimes should be stressed; $\bar\gamma=\gamma$ and $\bar\gamma > \gamma$. In the first 
regime, the system excitations are transferred to the environment exponentially. The system is approximately in a Markovian regime (it will 
only be exactly Markovian in the continuum limit) and almost no ``back action'' can be observed. In the second regime, the excitations 
oscillate between the main oscillator and the environment presenting a non-Markovian behavior. This indicates that, in this model, we can go 
from Markovian to a non-Markovian regime by only changing $\bar{\gamma}$, where the non-Markovian degree can be quantified in terms of this 
parameter. In the next sections, we will discuss quantum Darwinism in this model and its connection with non-Markovianity. 

\section{The Partial Information Plot approach to the model}\label{resultsPIP}

For both constant and non-constant coupling cases, we assume in this section the initial state of the main oscillator to be a ``cat state,'' 
where we set 
the parameters $a=1$, $b=1$, $\alpha_0=3$, so the global system state is given by
\begin{equation}
\ket{\Psi(0)}=G\left[ \left( \ket{\alpha_0} + \ket{-\alpha_0} \right) \otimes\prod_{k=1}^N\ket{0_k}\right]\noindent,
\label{catstate}
\end{equation}
with
\begin{equation}
G=\frac{1}{\left[  2\left(  1+e^{-2}   \right)  \right]^{1/2}}.
\end{equation}

It is straightforward to show that 
\begin{align}
|\braket{\alpha(t)|-\alpha(t)}|^2 &=e^{-2|\alpha(t)|^2}\label{prodalphas}\\
\prod_{k\in E}|\braket{\lambda_{k}(t)|-\lambda_{k}(t)}|^2 &=e^{-2(|\alpha_{0}|^{2}-|\alpha(t)|^2)},\label{prodlambdas}
\end{align}

On the other hand, $\alpha(t)$ is the solution of a linear differential equation with initial condition $\alpha_{0}$, therefore 
$|\alpha(t)|^2$ is proportional to $|\alpha_{0}|^2$ for all $t$.  Then, from Eqs.~\eqref{evolvedstate}, \eqref{prodalphas} and, \eqref{prodlambdas}, if $|\alpha_{0}|\gg 1$ and 
$\Gamma t\sim 1$, we see that states of the system and environment (as a whole) in Eq.~\eqref{evolvedstate} are approximately orthogonal, since both 
$\braket{\alpha(t)|-\alpha(t)}$ and $\braket{\Lambda(t)|-\Lambda(t)}$ are $\approx e^{-c|\alpha_{0}|^2}$ for some constant $c\sim 1$ 
independent of $|\alpha_{0}|^{2}$ (for this to be true in the nonconstant coupling case, one must avoid those specific instants of time where 
all excitations are in the environment). As mentioned in Sec.~\ref{QDT}, in the quantum Darwinistic regime the environment must monitor 
a preferential observable of the system. For the quantum Brownian motion studied in Ref.~\cite{zur2008}, it was essentially the oscillator 
position. In our model, with the initial state of the system defined in Eq.~\eqref{catstate}, it is already possible to infer the monitored 
observable. Indeed, from Eq.~\eqref{evolvedstate} and the above discussion, we see that 
our evolution closely matches that of the von Neumann measurement scheme [see Eq.~\eqref{exemplodarwinismo} and subsequent discussion]. 
Moreover, depending on the (complex) 
value of $\alpha_0$, the system can be monitored in position, momentum, or any quadrature. That is, the environment 
seems to be performing an approximate homodyne measurement. If $\alpha_0$ is a real number, the system will be in a superposition of states 
with 
distinct positions, leading the environment to monitor the position. But, if it is a pure imaginary number, the environment will monitor 
momentum, and so on.  

\subsection{Constant Coupling}\label{sconcoupl}

We assume now that $\gamma_k=0.1/30$ for all $k$ and the same distribution of frequencies as before. Recall that, as shown in Fig. 
\ref{conscoupl}, the excitations of the main oscillator decay exponentially, as expected from the analytical approximation. For large times, 
therefore, essentially all global system excitations can be found in the environment. 

For every fragment $F$, and every instant of time $t$, it is possible to compute the mutual information $I(S:F)$~\cite{bet2014}.  
We then estimate the PIP $\bar{I}(f)$, for each instant of time $t$, by sorting, for each value of $f$, $100$ 
fragments $F$ with $fN$ subsystems, and averaging the mutual information:  
\begin{equation}
\bar{I}(f)\approx\text{avg}_{(\text{random samples of }F)}I(S:F).
\end{equation}
Then, we repeated this for fragments composed from $1$ to $900$ ($100\%$) environment oscillators and varied $\Gamma t$ from $0$ to 
roughly $10$. We note, however, that the mutual information for each $F$ can be computed exactly, exploiting the fact that the reduced 
density matrix for $S+F$ has rank $2$ for all $F$ (see Ref.~\cite{bet2014} for details).

\begin{figure}
 \includegraphics[width=.7\linewidth]{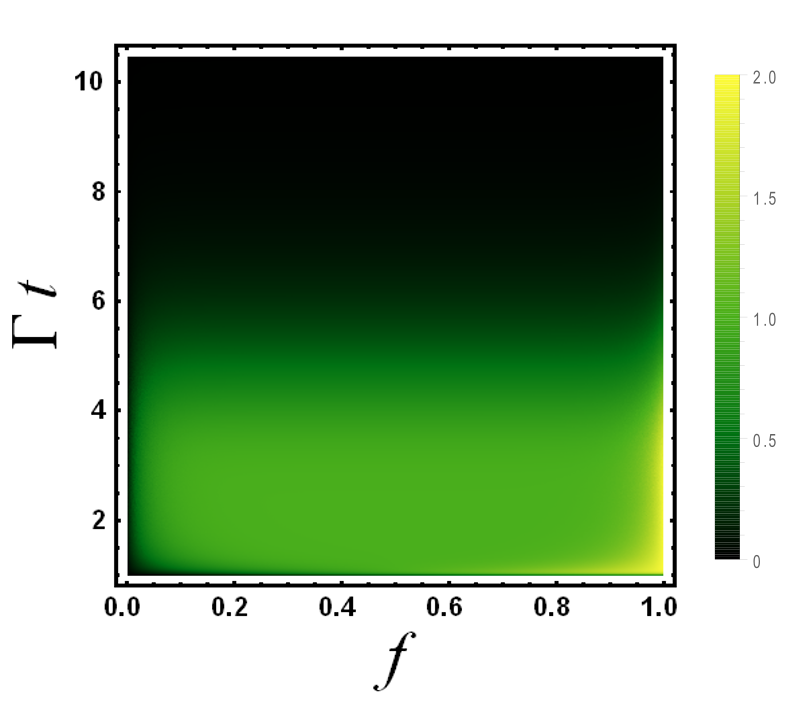}
\captionof{figure}{Average mutual information $\bar{I}(f)$ for a constant coupling with $\Gamma t$ going from $0$ to $10$ and the environment fractions $f$ from 
$1/900$ to $1$.}\label{consmapmutinf} 
 \end{figure}

The map in Fig. \ref{consmapmutinf} represents the averaged mutual information varying with time and with the environment 
fraction, $f$. We restrict our analysis in the rest of this section to time windows such that $\Gamma t\sim 6.0$ for all values of $f$. This is reasonable since after 
this time all system excitations are transferred to the environment and it decouples from the system. After a large enough time, 
the information about the system will be spread in the bath very redundantly. However the amount of information will be 
insignificant. Then, more specifically, we will analyze the mutual information, $f_{\delta}$ and $R_{\delta}$ until 
$\Gamma t=6.0$. 
 
\begin{figure}[h!]
 \includegraphics[width=.7\linewidth]{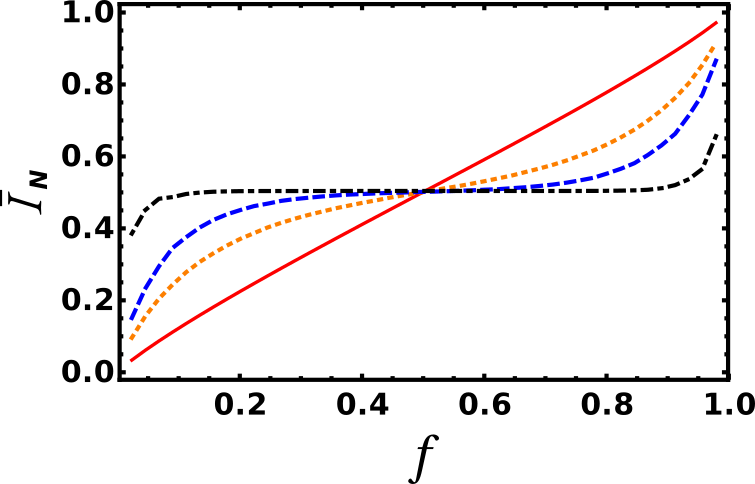}
\captionof{figure}{Normalized mutual information $\bar{I}_{N}(f)$ for the constant coupling case. The red (solid), orange (dotted), blue (dashed), and black 
(dot-dashed) lines represent the PIP for $\Gamma t=0.03$, $\Gamma t=0.17$, $\Gamma t=0.35$, and 
$\Gamma t=6$, respectively.}\label{mutinf0}
 \end{figure}

The signature of quantum Darwinism in the PIP approach is a plateau in the PIP. However, the maximal value of 
the mutual information can change with time. To compare the shape of each curve we normalize the mutual information for all time 
values; then, the normalized mutual information $\bar{I}_{N}(f)$ will range from $0$ to $1$. The plot in Fig. \ref{mutinf0} shows the PIP for some instants of time, from $\Gamma t=0.03$ to $\Gamma t=6$. 

For a plateau, the mutual information reaches a relevant value quickly and stabilizes until almost all the
environment is taken. In Fig. \ref{mutinf0}, for a small time ($\Gamma t=0,02$), the mutual information varies almost linearly with the size 
of the 
environment fragment (red line). This happens because, in the beginning, just a small amount of correlations were created between the 
environment and the 
main oscillator. For larger times, however, not only correlations become more significant, but information about the system is 
also spread redundantly on the oscillators of the environment. Indeed, in the black dashed dotted line in Fig. \ref{mutinf0}, the normalized mutual information, as a 
function of $f$, raises quickly and stabilizes. That is, a small fragment of the environment is enough to obtain almost all information about 
the preferential observable available in the environment. In this same figure, by the red (solid), orange (dotted), blue (dashed), and black 
(dot-dashed) lines, it is clear that as $\Gamma t$ grows, the plateau gets more and more pronounced.

\begin{center}
  \begin{figure}[h!]
	\flushleft
 \includegraphics[width=.7\linewidth]{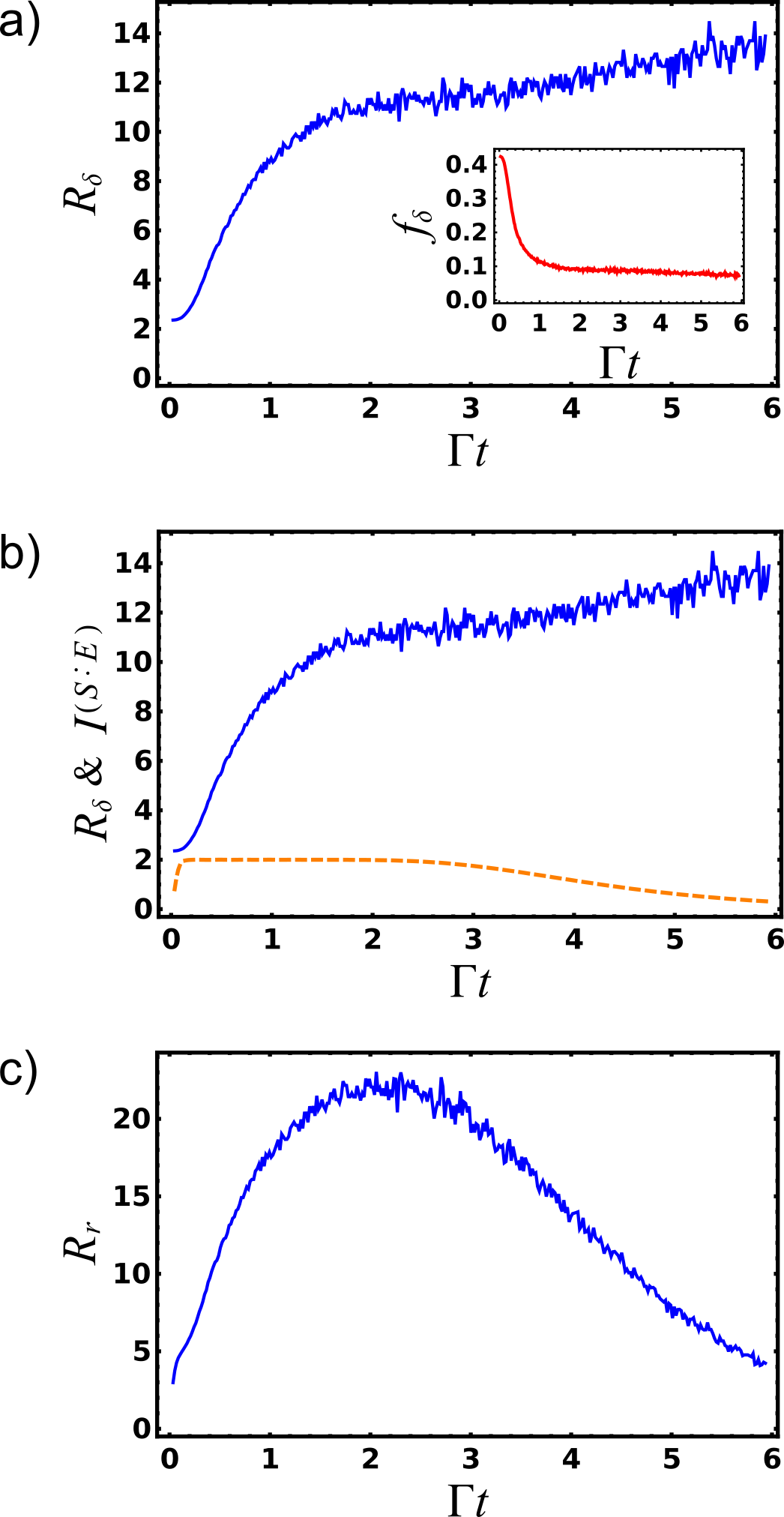}
\captionof{figure}{For $\delta=0.05$ and varying with time. (a) Redundancy $R_\delta$. The inset shows $f_\delta$. (b) Redundancy and mutual information between the system and the whole environment. (c) Relative redundancy. }\label{consfrdelta}
 \end{figure}
\end{center}

As discussed in the Sec.~\ref{QDT}, a figure of merit of quantum Darwinism is the fraction $f_\delta$ and the redundancy, 
$R_\delta$, [see Eq.~\eqref{rdelta}]. For constant coupling and $\delta=0.05$, we observed that the fraction $f_\delta$ decreases rapidly 
with time, [see inset in the Fig.~\ref{consfrdelta}(a)], while, of course, $R_\delta$ grows, see Fig.~\ref{consfrdelta}(a). However, as time 
evolves, although the PIP becomes more similar to a plateau, the total information contained in the environment 
about the system decreases substantially, see Fig.~\ref{consfrdelta}(b). 

We propose, then, to quantify the redundancy proportionally to the total amount of information about the main oscillator that is available in the environment, namely, $I(S:E)$. We introduce a new quantity, the \textit{relative redundancy}, 
$R_r$, given by
\begin{equation}
R_r(t)=R_\delta(t)I(S:E)(t),\noindent
\end{equation}
where $I(S:E)(t)$ is the total mutual information between the system and the whole environment at time $t$.

As shown in Fig. \ref{consfrdelta}(c), different from the redundancy, which grows monotonically, the relative redundancy grows with $\Gamma t$ for a maximal value but then decreases to zero.

\subsection{Non-Constant Coupling}

Starting from the same parameters of Sec.~\ref{sconcoupl} and the same initial cat state, Eq.\eqref{catstate}, we now consider the 
nonconstant coupling case, varying from a very small to a larger value of $\bar{\gamma}$.

\begin{figure}[h!]
\includegraphics[width=.7\linewidth]{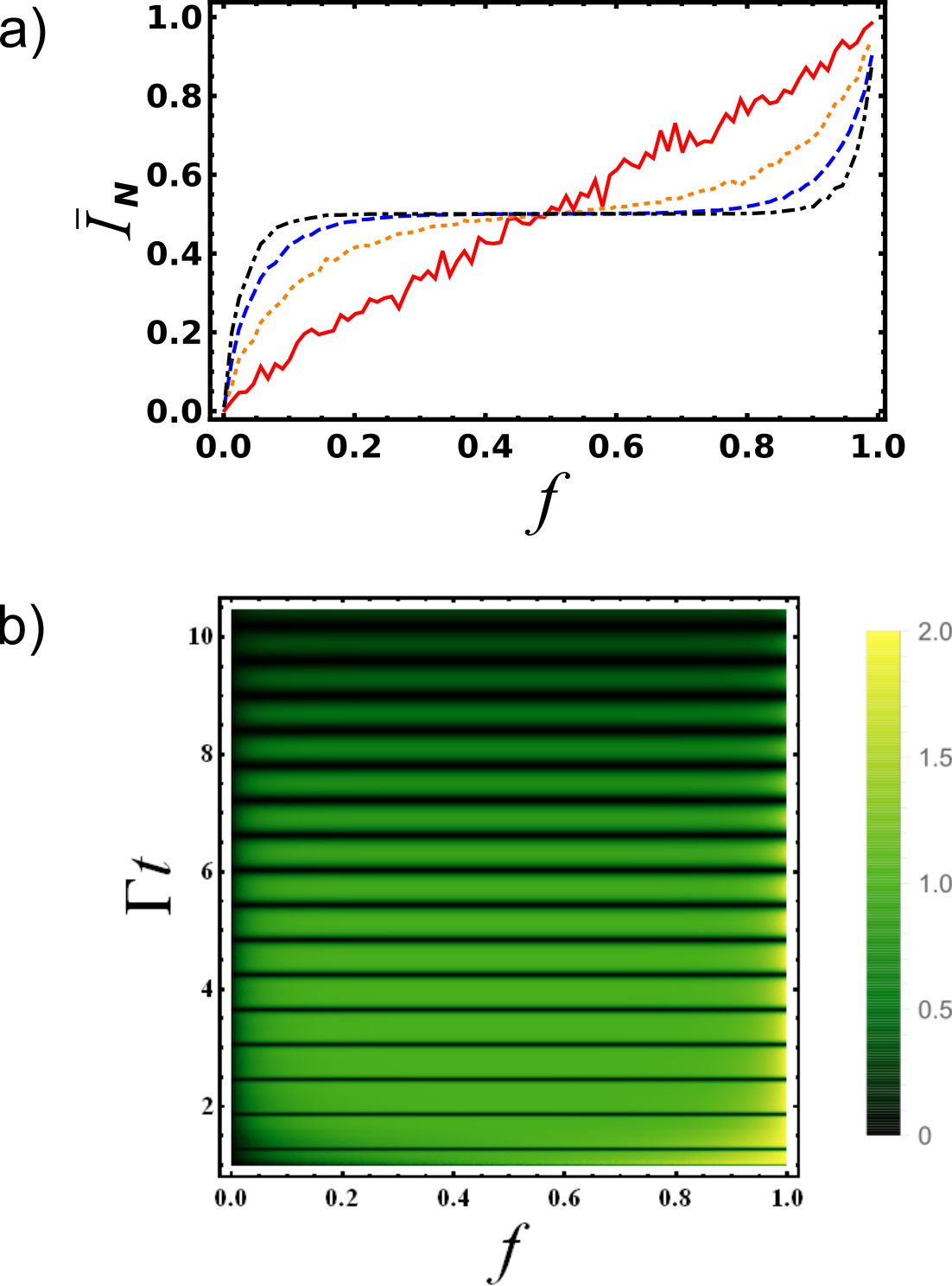}
\caption{For $\bar{\gamma}=50\gamma$: (a) Mutual Information versus $f$ for $\Gamma t=0.07$ (red, solid line), $\Gamma t=0.52$ 
(orange, dotted line), $\Gamma t=1.04$ (blue, dashed line) and $\Gamma t=6.0$ (black, dot-dashed line). (b) Map of the mutual information 
varying with $\Gamma t$ and with $f$.}\label{mapmutinf}
\end{figure}

For $\bar{\gamma}>\gamma$, we can see that the excitations oscillates between the main system and the environment. The magnitude of $\bar{\gamma}$ is related to the non-Markovianity of the system and these oscillations represent the backflow 
of information from the environment to the main system. In Fig. \ref{mapmutinf}(a), PIPs are shown 
for some instants of time. With $\bar{\gamma}=50\gamma$, at $\Gamma t=0.07$, (red, solid line), the mutual information 
depends almost linearly on the fragment size $f$. At $\Gamma t=0.52$, (orange, dotted line) and $\Gamma t=1.04$, (blue, dashed line) the 
plots are closer to a plateau. Lastly, at $\Gamma t=6.0$, (black, dot-dashed line), it is possible to see more clearly a plateau. Observing 
the strips on the map represented in Fig. \ref{mapmutinf}(b), we can see that this behavior repeats in each cycle. 

\begin{center}
\begin{figure}[h!]
 \includegraphics[width=.7\linewidth]{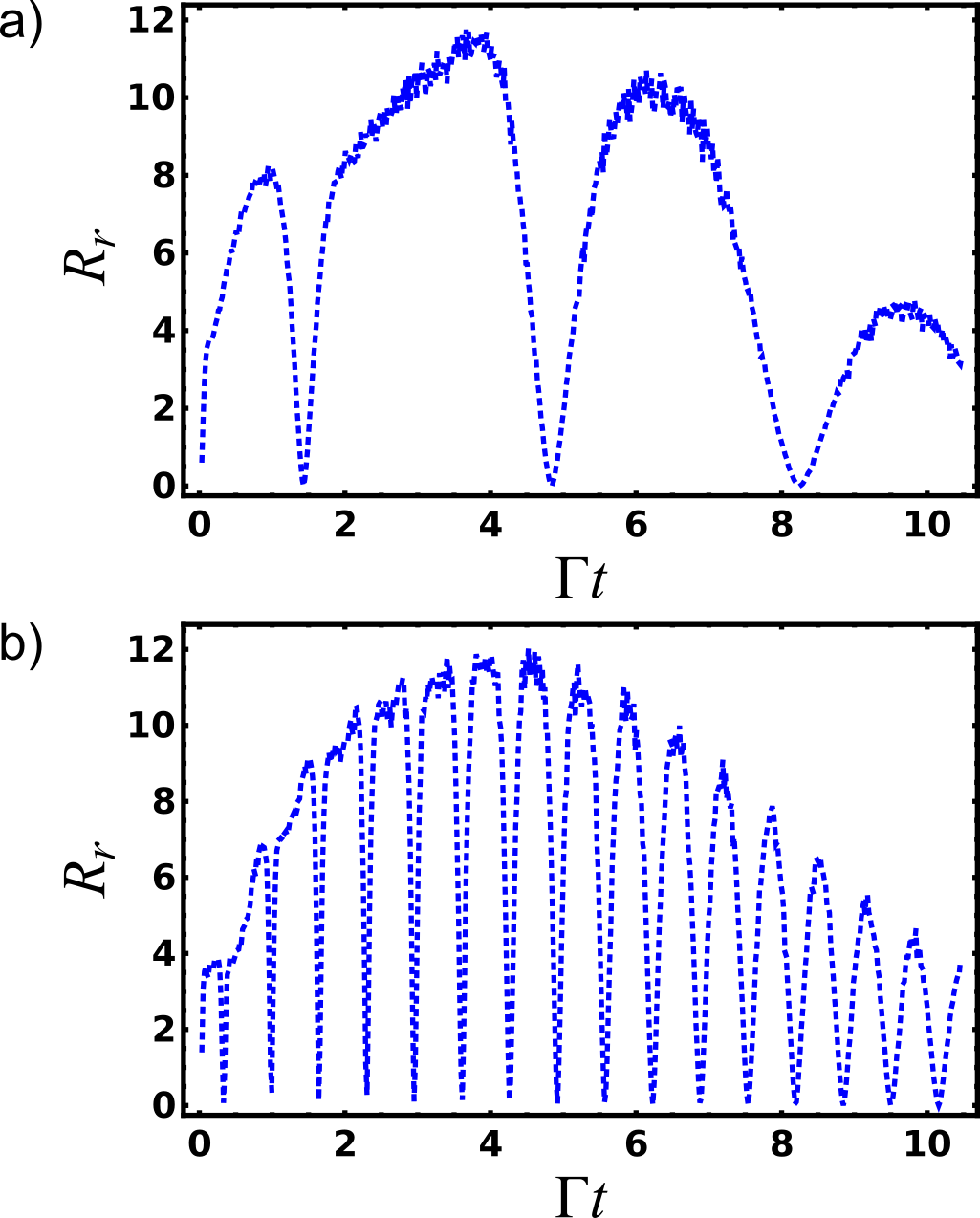}
\captionof{figure}{Relative redundancy for $\delta=0.05$ varying with $\Gamma t$ for 
(a) $\bar{\gamma}=10\gamma$ and (b) $\bar{\gamma}=50\gamma$.}\label{excimap}
\end{figure}
\end{center}

We have also computed the relative redundancy, $R_r$, as a function of time, in the same time window. In Figs.~\ref{excimap} =(a) and (b), it 
is possible to observe $R_r$ as a function of time for two values of $\bar\gamma$, with $\delta=0.05$. As the excitations exchange between the main system 
and environment oscillates in time (the orange dashed line), the redundancy and, consequently, the relative redundancy (the blue solid 
line) also oscillates with the same frequency. Consequently, the larger $\bar\gamma$ is, the larger is the oscillation frequency of $R_r$. 
In each cycle, as the excitations are transferred from the system to the environment, $R_r$ increases to a maximal value, and as the 
excitations flow back to the system, it decreases to a minimal value.

To sum up, in general there is significant redundant information in the environment, except in the vicinity of those instants of time where 
the oscillator loses all its excitations to the environment, since they also lose all correlations at those instants.

\section{The BPH approach to the model}\label{resultsBPH}

Let us consider again the initial state Eq.~\eqref{cat} with $a$ and $b$ arbitrary. From Eq.~\eqref{evolvedstate}, 
we can compute, for every $t$, the mapping from the initial main oscillator states 
$\rho=G^{2}(a\ket{\alpha_{0}}+b\ket{-\alpha_{0}})(a^{*}\bra{\alpha_{0}}+b^{*}\bra{-\alpha_{0}})$ to environment oscillator 
states:
\begin{align}
\Lambda_{F}^{t}(\rho)&=\text{Tr}_{S,E-F}(\ket{\Psi(t)}\bra{\Psi(t)})\\
&=G^2|a|^{2}\prod_{j\in F}\ket{\lambda_{j}(t)}\bra{\lambda_{j}(t)}\\
&+G^2|b|^2\prod_{j\in F}\ket{-\lambda_{j}(t)}\bra{-\lambda_{j}(t)}\\
&+D(t)\prod_{j\in F}\ket{\lambda_{j}(t)}\bra{-\lambda_{j}(t)}+\text{H.c.},
\end{align}
where
$$D(t)=G^2ab^{*}\braket{\alpha(t)|-\alpha(t)}\prod_{j\in E-F}\braket{\lambda_{j}(t)|-\lambda_{j}(t)}.$$
Now, assuming, as in the previous section, that $|\alpha_{0}|\gg 1$ we see that
$\braket{\alpha_{0}|-\alpha_{0}}=e^{-2|\alpha_{0}|^{2}}\approx 0$ and, therefore, $G\approx 1$. Moreover, we have $|a|^{2}\approx 
\text{Tr}(\ket{\alpha_{0}}\bra{\alpha_{0}}\rho)$ and $|b|^{2}\approx 
\text{Tr}(\ket{-\alpha_{0}}\bra{-\alpha_{0}}\rho)$. Assuming further that $D(t)\approx 0$, we have:
\begin{align}
 \Lambda_{F}^{t}(\rho)& \approx \text{Tr}(\ket{\alpha_{0}}\bra{\alpha_{0}}\rho)\prod_{j\in F}\ket{\lambda_{j}(t)}\bra{\lambda_{j}(t)}\\
&+\text{Tr}(\ket{-\alpha_{0}}\bra{-\alpha_{0}}\rho)\prod_{j\in F}\ket{-\lambda_{j}(t)}\bra{-\lambda_{j}(t)}.\label{approxmap}
\end{align}
Therefore, at least in the subspace generated by $\ket{\pm\alpha_{0}}$, the maps are approximately a measure and 
prepare map with the (approximate) POVM $\{\ket{\pm \alpha_{0}}\bra{\pm \alpha_{0}}\}$ and environment states 
$\sigma_{F,\pm}^{t}=\prod_{j\in F}\ket{\pm \lambda_{j}(t)}\bra{\pm \lambda_{j}(t)}$.

Now, let us check that indeed $D(t)\approx 0$ distinguishes two regimes: $\Gamma t \sim 1$ and $\Gamma t \gg 1$. For the first case, 
$D(t)$ is indeed small since we have $\braket{\alpha(t)|-\alpha(t)}=e^{-c|\alpha_{0}|^{2}}\approx 0$ for some 
$c\sim 1$. Note that for the nonconstant coupling case, not only must we have $\Gamma t \sim 1$ but also consider specific instants of 
time where the excitations go back to the main oscillator (see Fig.~\ref{lorcoupl50}). 

For $\Gamma 
t\gg 1$ (but not so large that recurrences can take place due to the finite number of oscillators in the environment), we can assume that all 
excitations are essentially in the environment. A typical environment fragment $F$ will have then $|\alpha_{0}|^{2}f$ excitations, where 
$f=|F|/N$ is the fraction of environment systems that this fragment has (see Appendix~\ref{sample}). Therefore, the complementary region 
$E-F$ will have $|\alpha_{0}|^{2}(1-f)$ excitations and we have $\prod_{j\in E-F}\braket{\lambda_{j}(t)|-\lambda_{j}(t)}\approx 
e^{-2|\alpha_{0}|^2(1-f)}\approx 0$ as long as $f$ is not too close to $1$. Indeed, it holds $D(t)\approx 0$.

Still, in the regime $\Gamma t\gg 1$, the distinguishability between the pure states $\sigma_{F,\pm}^{t}$, which is given by $1-\prod_{j\in 
F}|\braket{\lambda_{j}(t)|-\lambda_{j}(t)}|$ can then be approximated by $1-e^{-2|\alpha_{0}|^2f}$, and quickly approaches to $1$ 
with the environment fraction $f$ that is taken. Note that this argument applies for both the constant and non-constant coupling cases. In 
Appendix~\ref{sample} we argue in more detail that this distinguishability will depend essentially on $f$ and $|\alpha_{0}|^2$, but not on 
$\bar{\gamma}$.

To sum up, we see that the first regime, $\Gamma t\sim 1$, resembles the first example given in Sec.~\ref{examples}, while the asymptotic regime $\Gamma t\gg 1$ resembles the second one, since the global states have a similar structure in each instance, with the correspondences $\ket{0(1)}\leftrightarrow\ket{(-)\alpha(t)}$ and $\ket{0(1)}_{k}\leftrightarrow\ket{(-)\lambda_{k}(t)}$. For $\Gamma t\sim 1$ there are strong correlations between small environment fractions and the main oscillator which are sufficient to obtain significant information about the preferred observable at instant of time $t$ or its statistics in the initial state, so Darwinism is meaningful here from both the PIP and BPH perspectives.  For $\Gamma t\gg 1$, on the other hand, there are essentially no correlations between the environment and the main oscillator, so it is meaningless to address Darwinism from the PIP perspective. However, small environment fragments still keep a record of the statistics of the preferred observable in the main oscillator initial state. So it is still meaningful to address Darwinism from the BPH perspective.

\section{Quantum Darwinism and non-Markovianity}\label{darwinismxmarkovianity}

In the constant coupling case, there are no excitations returning to the system, and, consequently, no backflow of information, signalizing 
that the system is Markovian. In the non-constant coupling case, the oscillations of the excitations of the system and the environment show 
us that these systems are non-Markovian. In particular, it seems that the non-Markovianity of the system raises with the frequency of these 
oscillations.

\begin{figure}[h!]
\includegraphics[width=.7\linewidth]{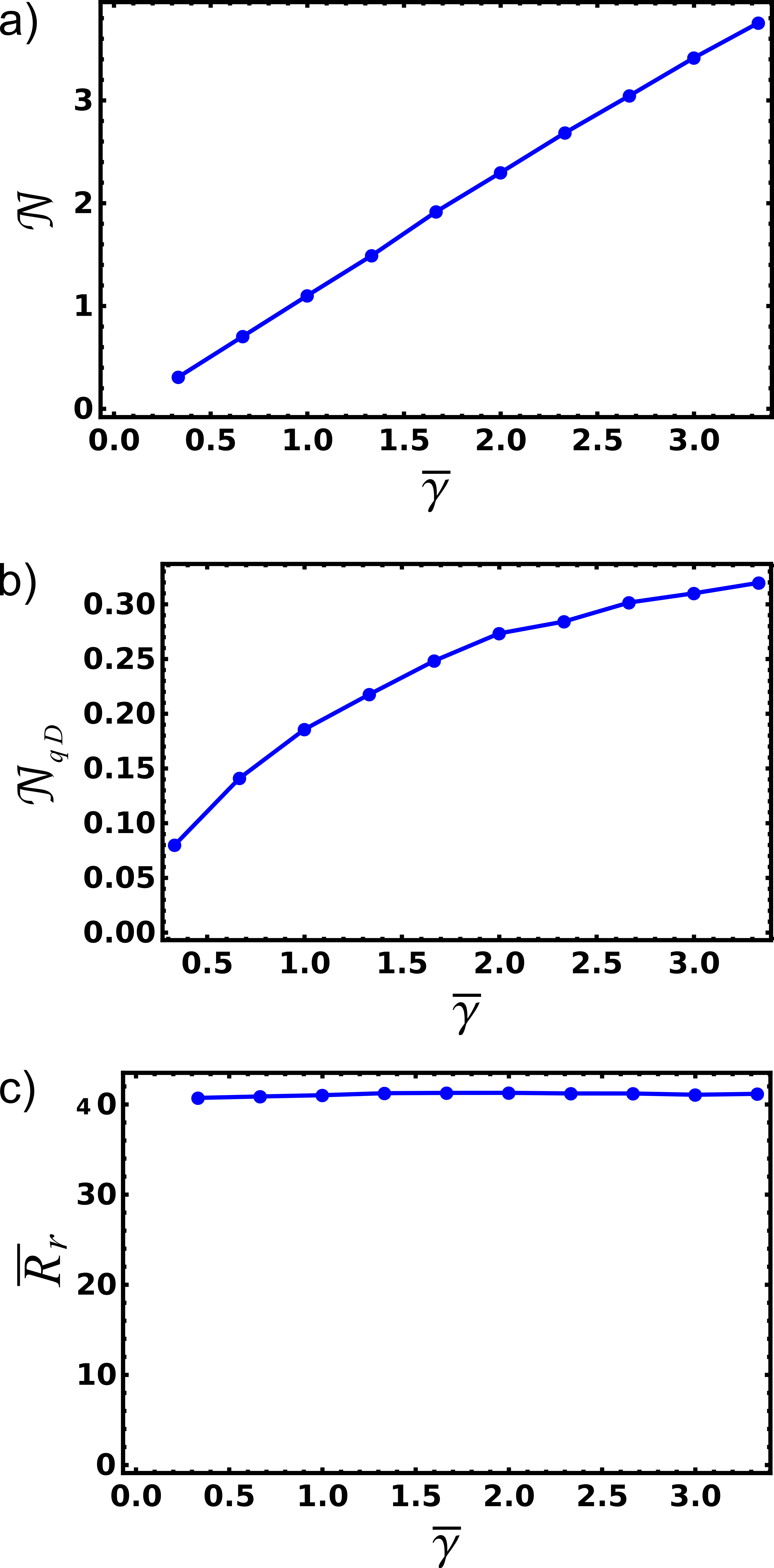}
\captionof{figure}{(a) non-Markovianity degree $\mathcal{N}$, (b) Non-monotonicity $\mathcal{N}_{qD}$, and (c) averaged relative redundancy $R_r$ versus $\bar{\gamma}$, with $\bar{\gamma}$ going from $10\gamma$ to $100\gamma$. In (a), we sort 
1000 initial states coherent states and $\mathcal{N}$ was calculated for all combinations of pairs and in (c) the time interval of the integration is $\Gamma\Delta t=10$.}\label{NMdegree}
 \end{figure}

We recall the non-Markovianity quantifier in Eq.~\eqref{deg}. We estimated this quantity by sorting pairs of initial coherent 
states for the main oscillator. We chose $1000$ initial complex values of $\alpha_0$, according to a normal distribution, and calculated the 
fidelity of all possible combinations of these states, for different values of $\bar{\gamma}$. As expected, the larger $\bar{\gamma}$ is the 
larger the non-Markovianity degree will be, see Fig. \ref{NMdegree}(a). 

To understand the connection between quantum Darwinism and non-Markovianity, in Ref.~\cite{sab2016} the authors quantified the 
\textit{non-monotonic behavior} of $f_\delta$ that was defined similarly to the non-Markovianity degree:
\begin{equation}
\mathcal{N}_{qD}=\int_{df_{\delta}/dt>0}\frac{df_{\delta}}{dt}dt.\noindent
\label{nonmono}
\end{equation}
Note that the integral is calculated just when $df_{\delta}/dt$ is positive. They show that the behavior of $\mathcal{N}_{qD}$ is very similar to the 
behavior of $\mathcal{N}$, having a good qualitative agreement. So, they concluded that quantum Darwinism is being hindered by 
non-Markovianity. 

We also calculated this quantity in our model and we could observe that $\mathcal{N}_{dD}$ as a function of $\bar{\gamma}$ grows 
monotonically as $\mathcal{N}$, see Fig.~\ref{NMdegree}(b). This is expected since as non-Markovianity grows, more oscillations in the 
redundancy and $f_\delta$ will be present. Then, if we integrate the values of $f_\delta$ just when it is growing, the non-monotonicity 
will increase with the oscillation frequency and therefore, with $\bar\gamma$.

Although this analogy is consistent, we present here a different point of view. We could observe in non-Markovian systems that the 
oscillations of the excitations lead to oscillations in the redundancy. When the excitations are being transferred to the environment, 
the redundancy grows. When the excitations are flowing back to the system the redundancy decreases. Nevertheless, even in the 
moments of backflow of excitations, we can say that there exists some redundancy in the system and, therefore, some degree of quantum 
Darwinism. We believe then that a more sensitive figure of merit would be the \emph{averaged} relative redundancy 
$R_r(\bar{\gamma})$ for a 
certain period of time $\Delta t = t\text{max}-t\text{min}$, that is,
\begin{equation}
\bar{R}_r(\bar{\gamma})=\frac{1}{\Delta t}\int_{t\text{min}}^{t\text{max}}R_r(t,\bar{\gamma}) dt.\noindent
\label{AR}
\end{equation}

We calculated this quantity for different values of $\bar{\gamma}$ in a time window $\Gamma\Delta t=10$ . The result is presented in 
Fig.~\ref{NMdegree}(c) and it show us that the values of $\bar{R}_r(\bar{\gamma})$ are practically 
constant. This implies that, for any random instant of time, the probability to find quantum Darwinism in the Markovian case is nearly the same of the case of high non-Markovianity.
Putting in another way, even if the amount of quantum 
Darwinism is non-monotonic in time in a non-Markovian dynamics, \emph{on average}, it is the same as in the Markovian case. 

We can also relate the two concepts from the BHP perspective. As discussed in Sec.~\ref{model}, in the asymptotic regime $\Gamma 
t\gg 1$, all excitations initially in the main oscillator go to the environment. Their distribution in the environment, however, 
depends strongly on $\bar{\gamma}$. Since the non-Markovianity of the main oscillator dynamics depends only on $\bar{\gamma}$, the 
asymptotic state of the environment actually keeps a record of the non-Markovianity. On the other hand, as pointed out in 
Sec.~\ref{resultsBPH}, and detailed in Appendix~\ref{sample}, the asymptotic environment state redundant recording of the main oscillator 
preferred observable (the quantum Darwinism from the BHP perspective) is essentially \emph{independent} of $\bar{\gamma}$.  

\section{Conclusions}\label{conclusion}

We studied quantum Darwinism from two distinct perspectives and its connection with non-Markovianity in a system made of a single quantum 
harmonic 
oscillator coupled to an environment of a large set of quantum harmonic oscillators. Through the calculations of mutual information between 
the system and fragments of the environment with different sizes, we could see that the environment monitors the system constantly and we 
could check the existence or not of quantum Darwinism at each instant of time. We analyzed the mutual 
information varying with the environment fragment size and the relative redundancy. This was done also varying a parameter of the model 
that controls the degree of non-Markovianity of the main oscillator dynamics.

In the Markovian case, we verified that, after correlations are established between the system and the environment, indeed the 
information about the preferred observable is distributed redundantly through the environment. However, in the non-Markovian case, the 
excitations and the relative redundancy oscillates. As the non-Markovianity increases, more oscillations 
takes place. In both cases, however, it is noticeable that the environment monitors the system instantly.

 In Ref.~\cite{sab2016},  the authors study the consequences of non-Markovianity on quantum Darwinism through the 
non-monotonicity of $f_\delta$, quantified by $\mathcal{N}_f$. There, the value of $f_\delta$ is integrated just when  
$\dot{f}_\delta$ is growing. It was shown that this quantity increases with the non-Markovianity degree, 
suggesting that quantum Darwinism is being hindered by it. We could observe a 
similar result in our model. 

We offer in this paper a different point of view to relate non-Markovianity and quantum 
Darwinism. In the time instants where the excitations are being transferred from the system to the environment, correlations between them are created, spreading redundant information about the system in the environment. This also causes a rise in the redundancy. As 
expected in a non-Markovian system, the oscillations will cause a backflow of information from the environment to the system. When the 
environment ``gives back'' part of the information for the system, a part of the correlation early created is also destroyed, leading to a 
decrease of the redundancy. In these moments, it is also possible to see the relative redundancy decreasing. However, as the transfer of 
excitations oscillates, the correlations turn to grow up and this cycle repeats. Then, since quantum Darwinism can be observed in 
cycles, even when it is decreasing, it makes sense to consider it in all instants of time.  Then, to quantify the effect of 
non-Markovianity on quantum Darwinism, we computed the \emph{averaged} relative redundancy $R_r$, that consists 
of the relative redundancy averaged on a period of time, for different values of $\bar\gamma$. We observed that there is no clear 
correlation between $R_r$ and $\bar\gamma$, leading us to conclude that as long the system and 
environment are exchanging excitations, the observation of the quantum Darwinism depends just on the particular instant of time of the analysis.

Finally, even in the asymptotic limit, it is possible to address quantum Darwinism from the BPH perspective, even though 
there is no correlations between the main oscillator and its environment. It is also possible to address non-Markovianity in this limit, 
since the distribution of excitations in the environment is sensitive to the non-Markovianity of the main oscillator dynamics. Then, we 
have seen that, also from this perspective, that quantum Darwinism in this regime 
is essentially independent from the non-Markovianity degree.   

\acknowledgments
This work was supported by the Conselho Nacional de Desenvolvimento Cient\'ifico e Tecnol\'ogico (CNPq) and the Coordena\c{c}\~ao de 
Aperfei\c{c}oamento de Pessoal de N\'ivel Superior (CAPES). We thank Sabrina Maniscalco for fruitiful discussions and the Okinawa 
Institute of Science and Technology (OIST), which made possible the meeting of S.M.O. with Professor Maniscalco.

\bibliography{ref}

\begin{thebibliography}{10}

\bibitem{schcat}
E.~Schr{\"o}dinger, {\em The interpretation of quantum mechanics: Dublin
  seminars (1949-1955) and other unpublished essays}.
\newblock Ox Bow Pr, 1995.

\bibitem{epr}
A.~Einstein, B.~Podolsky, and N.~Rosen, ``Can quantum-mechanical description of
  physical reality be considered complete?,'' {\em Physical review}, vol.~47,
  no.~10, p.~777, 1935.

\bibitem{zur1981}
W.~H. Zurek, ``Pointer basis of quantum apparatus: Into what mixture does the
  wave packet collapse?,'' {\em Physical review D}, vol.~24, no.~6, p.~1516,
  1981.

\bibitem{zur2003}
W.~H. Zurek, ``Decoherence, einselection, and the quantum origins of the
  classical,'' {\em Reviews of modern physics}, vol.~75, no.~3, p.~715, 2003.

\bibitem{schl2007}
M.~A. Schlosshauer, {\em Decoherence: and the quantum-to-classical transition}.
\newblock Springer Science \& Business Media, 2007.

\bibitem{zur1982}
W.~H. Zurek, ``Environment-induced superselection rules,'' {\em Physical Review
  D}, vol.~26, no.~8, p.~1862, 1982.

\bibitem{bru1996}
M.~Brune, E.~Hagley, J.~Dreyer, X.~Maitre, A.~Maali, C.~Wunderlich, J.~Raimond,
  and S.~Haroche, ``Observing the progressive decoherence of the “meter” in
  a quantum measurement,'' {\em Physical Review Letters}, vol.~77, no.~24,
  p.~4887, 1996.

\bibitem{dav1996}
L.~Davidovich, M.~Brune, J.~Raimond, and S.~Haroche, ``Mesoscopic quantum
  coherences in cavity qed: Preparation and decoherence monitoring schemes,''
  {\em Physical Review A}, vol.~53, no.~3, p.~1295, 1996.

\bibitem{bre2002}
H.-P. Breuer and F.~Petruccione, {\em The theory of open quantum systems}.
\newblock Oxford University Press on Demand, 2002.

\bibitem{zur2005a}
F.~Cucchietti, J.~P. Paz, and W.~Zurek, ``Decoherence from spin environments,''
  {\em Physical Review A}, vol.~72, no.~5, p.~052113, 2005.

\bibitem{zur2004}
H.~Ollivier, D.~Poulin, and W.~H. Zurek, ``Objective properties from subjective
  quantum states: Environment as a witness,'' {\em Physical review letters},
  vol.~93, no.~22, p.~220401, 2004.

\bibitem{zur2003a}
W.~H. Zurek, ``Quantum darwinism and envariance,'' {\em arXiv preprint
  quant-ph/0308163}, 2003.

\bibitem{zur2005}
R.~Blume-Kohout and W.~H. Zurek, ``A simple example of “quantum darwinism”:
  Redundant information storage in many-spin environments,'' {\em Foundations
  of Physics}, vol.~35, no.~11, pp.~1857--1876, 2005.

\bibitem{zur2006}
R.~Blume-Kohout and W.~H. Zurek, ``Quantum darwinism: Entanglement, branches,
  and the emergent classicality of redundantly stored quantum information,''
  {\em Physical Review A}, vol.~73, no.~6, p.~062310, 2006.

\bibitem{zur2005b}
H.~Ollivier, D.~Poulin, and W.~H. Zurek, ``Environment as a witness: Selective
  proliferation of information and emergence of objectivity in a quantum
  universe,'' {\em Physical review A}, vol.~72, no.~4, p.~042113, 2005.

\bibitem{zunatk2009}
W.~H. Zurek, ``Quantum darwinism,'' {\em Nature Physics}, vol.~5, no.~3,
  p.~181, 2009.

\bibitem{zur2009}
M.~Zwolak, H.~Quan, and W.~H. Zurek, ``Quantum darwinism in a mixed
  environment,'' {\em Physical review letters}, vol.~103, no.~11, p.~110402,
  2009.

\bibitem{zur2008}
R.~Blume-Kohout and W.~H. Zurek, ``Quantum darwinism in quantum brownian
  motion,'' {\em Physical review letters}, vol.~101, no.~24, p.~240405, 2008.

\bibitem{zur2018}
T.~Unden, D.~Louzon, M.~Zwolak, W.~Zurek, and F.~Jelezko, ``Revealing the
  emergence of classicality in nitrogen-vacancy centers,'' {\em arXiv preprint
  arXiv:1809.10456}, 2018.

\bibitem{mas2019}
M.~P. Nadia~milazzo, Salvatore~Lorenzo and G.~M. Palma, ``The role of
  information back flow in the emergence of quantum darwinism,'' {\em arXiv
  preprint arXiv:1901.05826}, 2019.

\bibitem{sab2016}
F.~Galve, R.~Zambrini, and S.~Maniscalco, ``Non-markovianity hinders quantum
  darwinism,'' {\em Scientific reports}, vol.~6, p.~19607, 2016.

\bibitem{bra2015}
F.~G. Brand{\~a}o, M.~Piani, and P.~Horodecki, ``Generic emergence of classical
  features in quantum darwinism,'' {\em Nature communications}, vol.~6,
  p.~7908, 2015.

\bibitem{hor2015}
R.~Horodecki, J.~K. Korbicz, and P.~Horodecki, ``Quantum origins of
  objectivity,'' {\em Phys. Rev. A}, vol.~91, p.~032122, Mar 2015.

\bibitem{von2018mathematical}
J.~Von~Neumann, {\em Mathematical foundations of quantum mechanics: New
  edition}.
\newblock Princeton university press, 2018.

\bibitem{kno2018}
P.~A. Knott, T.~Tufarelli, M.~Piani, and G.~Adesso, ``Generic emergence of
  objectivity of observables in infinite dimensions,'' {\em arXiv preprint
  arXiv:1802.05719}, 2018.

\bibitem{wol2008}
M.~M. Wolf and J.~I. Cirac, ``Dividing quantum channels,'' {\em Communications
  in Mathematical Physics}, vol.~279, no.~1, pp.~147--168, 2008.

\bibitem{riv2014}
A.~Rivas, S.~F. Huelga, and M.~B. Plenio, ``Quantum non-markovianity:
  characterization, quantification and detection,'' {\em Reports on Progress in
  Physics}, vol.~77, no.~9, p.~094001, 2014.

\bibitem{nad2015}
N.~K. Bernardes, A.~Cuevas, A.~Orieux, C.~Monken, P.~Mataloni, F.~Sciarrino,
  and M.~F. Santos, ``Experimental observation of weak non-markovianity,'' {\em
  Scientific reports}, vol.~5, p.~17520, 2015.

\bibitem{nad2017}
N.~K. Bernardes, A.~R. Carvalho, C.~H. Monken, and M.~F. Santos, ``Coarse
  graining a non-markovian collisional model,'' {\em Physical Review A},
  vol.~95, no.~3, p.~032117, 2017.

\bibitem{lac2015}
D.~Lacroix, V.~Sargsyan, G.~Adamian, and N.~Antonenko, ``Description of
  non-markovian effect in open quantum system with the discretized environment
  method,'' {\em The European Physical Journal B}, vol.~88, no.~4, p.~89, 2015.

\bibitem{sab2011}
R.~Vasile, S.~Maniscalco, M.~G. Paris, H.-P. Breuer, and J.~Piilo,
  ``Quantifying non-markovianity of continuous-variable gaussian dynamical
  maps,'' {\em Physical Review A}, vol.~84, no.~5, p.~052118, 2011.

\bibitem{sab2017}
M.~Cianciaruso, S.~Maniscalco, and G.~Adesso, ``Role of non-markovianity and
  backflow of information in the speed of quantum evolution,'' {\em Physical
  Review A}, vol.~96, no.~1, p.~012105, 2017.

\bibitem{bre2009}
H.-P. Breuer, E.-M. Laine, and J.~Piilo, ``Measure for the degree of
  non-markovian behavior of quantum processes in open systems,'' {\em Physical
  review letters}, vol.~103, no.~21, p.~210401, 2009.

\bibitem{scu1997}
M.~O. Scully and M.~S. Zubairy, {\em Quantum optics}.
\newblock Cambridge university press, 1997.

\bibitem{bet2014}
A.~de~Paula~Jr, J.~de~Oliveira~Jr, J.~P. de~Faria, D.~S. Freitas, and M.~Nemes,
  ``Entanglement dynamics of many-body systems: Analytical results,'' {\em
  Physical Review A}, vol.~89, no.~2, p.~022303, 2014.

\bibitem{rap2018}
J.~P. Santos, A.~L. de~Paula~Jr, R.~Drumond, G.~T. Landi, and M.~Paternostro,
  ``Irreversibility at zero temperature from the perspective of the
  environment,'' {\em Physical Review A}, vol.~97, no.~5, p.~050101, 2018.

\end{thebibliography}

\appendix

\section{Exact solution through a continuum limit approximation}\label{exato}

As mentioned in section~\ref{model}, the model admits an analytical for the continuum limit Eq.~\eqref{contlimit}. Then, one is able to solve Eqs.~\eqref{alphadiff} and Eq.\eqref{lamdiff} by formally integrating \eqref{lamdiff}, substituting in Eq.~\eqref{alphadiff}, leading to an integro-differential equation that is easy to solve in the continuum limit. The result is then:
\begin{eqnarray}
\alpha(t)&=&\alpha_0e^{-\Gamma t/2 }\label{soluapprox},\\
\lambda(\omega,t)&=&i\alpha_0\gamma\frac{e^{\left(-\Gamma/2+i\Delta\omega\right)t}}{\Gamma/2 + i\Delta\omega},\noindent
\end{eqnarray}
where $\Delta\omega=\omega_0 - \omega$ and $\Gamma=4\pi \gamma^2\rho$ [recall Eq.~\eqref{defdec}]. 

In the asymptotic limit, the environment excitations distribute like a Lorenztian centered at $\omega_0$:
\begin{equation}
 |\lambda(\omega,t\rightarrow\infty)|^2\rho=|\alpha_0|^2\frac{1}{\pi}\frac{\Gamma}{(\Gamma/2)^2 + (\omega-\omega_0)^2}\label{lorentzmarkov}.
\end{equation}

For the nonconstant coupling case, by a similar procedure, one arrives at:
\begin{eqnarray}
&&\alpha(t)=\frac{\alpha_0}{2\Omega}e^{-\Gamma t/4 }\left[ \Gamma/4\left( e^{-\Omega t} - e^{\Omega t} \right) 
\right.\nonumber\\
&&+ \left.\Omega\left( e^{-\Omega t} + e^{\Omega t} \right) \right],\label{alphas}\\
&&\lambda(\omega_0,t)=\frac{i\alpha_0\bar{\gamma}}{2\Omega}e^{-\Gamma t/4}\left( e^{-\Omega t} - e^{\Omega t} \right),\label{lambda0s}\\
&&\lambda(\omega,t)=\frac{-i\alpha_0\gamma}{2\Omega}\left[ \frac{\left(\Gamma/4+\Omega\right)\left(   
1-e^{-\left(\Gamma/4+ i\Delta\omega+\Omega\right)t}  \right)}{\Gamma/4+ i\Delta\omega+\Omega}  \right.\nonumber\\
 &&+ \left. \frac{\left(\Gamma/4-\Omega\right)\left(   1-e^{-\left(\Gamma/4+ i\Delta\omega-\Omega\right)t}  
\right)}{-\Gamma/4- i\Delta\omega+\Omega}\right],\label{lambdas}\noindent
\end{eqnarray}
with $\Omega=\sqrt{(\Gamma/4)^2 - \bar{\gamma}^2}$. From these expressions we get that, in the asymptotic limit and for $\bar{\gamma}\gg \Gamma$, the environment excitations distribute approximately as a sum of two Lorentzians centered at $\omega_0+\bar{\gamma}$ and $\omega_0-\bar{\gamma}$:
\begin{align}
|\lambda(\omega,t\rightarrow\infty)|^2\rho&\approx \frac{|\alpha_0|^2}{4\pi}[\frac{\Gamma}{(\Gamma/4)^2 + (\omega-(\omega_0+\bar{\gamma}))^2}\nonumber \\
&+\frac{\Gamma}{(\Gamma/4)^2 + (\omega-(\omega_0-\bar{\gamma}))^2}].\label{lorentznonmarkov}
\end{align}

%\begin{eqnarray}
% &&|\lambda(\omega,t\rightarrow\infty)|^2\rho=|\alpha_0|^2 %%\frac{\Gamma}{\pi}\times\nonumber \\
% &&\left[\frac{(\Gamma/4)^2+(\Delta \omega)^2}{[(\Gamma/4)^2+%%(\omega-(\omega_0+\Omega))^2][(\Gamma/4)^2+(\omega-(\omega_0-\Omega))^2]}\right]. \nonumber \\ \label{lorentznonmarkov}
%\end{eqnarray}

% We see from Eq.~\eqref{soluapprox} that the energy of they main oscillator 
% dissipates at rate $2\Gamma$ (since it is proportional to squared modulus of $\alpha(t)$). On the other hand, in the regime that 
% $\bar{\gamma}\gg \Gamma$ Eq.~\eqref{alphas} shows that the main oscillator will dissipate at rate $\Gamma$. Loosely speaking, this is due 
% to the fact that the initial excitations in the main oscillator will spend roughly half of the time in the resonant oscillator, which, by 
% itself, is not subjected to dissipation. 

\section{Distinguishability of states of sampled environment subsystems}\label{sample}

We have discussed the analytical solution to the model in the continuum limit in Sec.~\ref{model} and Appendix~\ref{exato}. We recall that there is a well-defined 
asymptotic distribution of excitations in the reservoir, that is $\lim_{t\rightarrow\infty}|\lambda(\omega,t)|^2\rho$ exists for all $\omega$. 
For 
$\bar{\gamma}=\gamma$, the Markovian regime, we have a Lorentzian 
distribution of width $\Gamma$ [see Eq.~\eqref{lorentzmarkov}], while for $\bar{\gamma}\gg \Gamma$, deep in the non-Markovian 
regime, we 
have essentially a sum of two Lorentzians of width $\Gamma/2$ [see eq.~\eqref{lorentznonmarkov}]. Even though we consider a model with a \emph{finite} number of oscillators 
in the environment, this analytical solution serves as guide and a good approximation to understand the finite case.

We have seen in Sec.~\ref{resultsBPH} that, for an initial global state 
$G(a\ket{\alpha_{0}}+b\ket{-\alpha_{0}})\otimes \prod_{k}\ket{0}_{k}$, the mapping between the initial system state and the 
state of environment fragment $F$ at instant $t$ satisfies Eq.~\eqref{approxmap}, which we reproduce here for convenience:

\begin{align}
 &\Lambda_{F}^{t}(\rho) \approx \text{Tr}(\ket{\alpha_{0}}\bra{\alpha_{0}}\rho)\prod_{k\in 
F}\ket{\lambda_{k}(t)}\bra{\lambda_{k}(t)}\\
&+\text{Tr}(\ket{-\alpha_{0}}\bra{-\alpha_{0}}\rho)\prod_{k\in F}\ket{-\lambda_{k}(t)}\bra{-\lambda_{k}(t)}.
\end{align} 

The distinguishability between the environment states appearing in these expressions depends essentially on the number of excitations in the 
environment fraction $F$, that is, $\sum_{k\in F}|\lambda_{k}(t)|^2$.  We shall consider $F\subset E$, with $|F|=f/N$ for fixed $f>0$, 
as elements of a sample space with uniform probability distribution, with each $F$ having probability $1/\binom{N}{fN}$, where 
$\binom{N}{fN}$ is the binomial coefficient. We would like to understand a random variable of the form

\begin{equation}
 X(F)=\sum_{k\in F}p_{k},
\end{equation}
where $p_{k}=|\lambda_{k}(T)|^2$ for some fixed (independent of $N$) $T\gg 0$. It holds that $\sum_{i=1}^{N}p_{k}=|\alpha_{0}|^2:=p$ holds exactly for all $T$ and the constant $p$ is 
independent of $N$. 

Basically, we want to show the distribution of $X$ to be highly concentrated around its average value, if $N$ is large enough. 

It is easy to check that the expectation value of $X$ satisfies
\begin{equation}
 \mathbb{E}(X)=\frac{1}{\binom{N}{fN}}\sum_{F\subset E}\sum_{k\in F}p_{k}=pf,
\end{equation}
regardless of the functional dependence of $p_{k}$ on $k$.

We shall consider first a a prototype distribution for the distribution we have in the Markovian case 
[which is the Lorentzian Eq.~\eqref{lorentzmarkov} in the continuum limit].  Namely, for some fixed $0<a<1$, $p_{k}=p/Na$ for 
$|k/N-1/2|<a/2$ and $p_{k}=0$ otherwise. That is, 
the graph of $p_{k}$  has a rectangular shape of width $a$, height $p/Na$, centered at $N/2$. 
Let $[N]=\{1,...,N\}$ and let $J=\{k\in [N]||k/N-1/2|<a/2\}$, that is, $J$ is the set of indexes where $p_{k}\neq 0$. The value of $X(F)$ will then be defined by $|F\cap J|$, that is, the number of indexes of $F$ that 
are in $J$. Now, if we think the points of $F$ as ``particles'' that we can uniformly distribute in a large ``box'' $[N]$, that was 
divided in two parts, a smaller box $J$ versus the rest $[N]-J$, we have a standard problem in statistical mechanics. It is well known that the density of particles in $J$, $|F\cap J|/aN$ will be highly concentrated, in probability, around its average value 
$f$. Namely, for any $\epsilon>0$, it holds that, for sufficiently large $N$,

\begin{equation}
 \mathbb{P}(|\frac{|F\cap J|}{aN}-f|)>\epsilon\leq be^{-cN}
\end{equation}
for some positive constants $b$ and $c$ that depends on $f$ and $a$. Since 
$X(F)=\frac{p}{aN}|F\cap J|$, we then get:
\begin{equation}
 \mathbb{P}(|X-pf|)>p\epsilon\leq be^{-cN}.\label{concentrationbound}
\end{equation}

That is, for the vast majority of environment fractions $F$ with $fN$ subsystems, the total number of excitations in those fractions will 
be very close to $f|\alpha_{0}|^2$. 

We can also consider a prototype distribution in the same spirit, but for the non-Markovian 
case [a sum of two Lorentzians, in the continuum limit and for $\bar{\gamma} \gg \Gamma$, see Eq.~\eqref{lorentznonmarkov}]. Namely, for some fixed $0<a<1$, 
$p_{k}=p/Na$ for $|k/N-1/4|<a/4$ or $|k/N-3/4|<a/4$, and $p_{k}=0$ otherwise. In this case, the 
graph of $p_{k}$ has the shape of two rectangles of width $a/2$, height $p/Na$, one centered at $N/4$, 
the other at $3N/4$. Nevertheless, the exact same reasoning as before applies and we get exactly the same bound Eq.~\eqref{concentrationbound}.

\end{document}